\documentclass[prd,onecolumn,a4paper,nofootinbib,12pt,eqsecnum]{revtex4-1}
\pdfoutput=1
\usepackage{textcase}
\usepackage{bm}
\usepackage{amsfonts}
\usepackage{lipsum}
\usepackage{amsthm}
\usepackage{setspace}
\usepackage{hyperref}
\usepackage{slashed}
\usepackage{colortbl}
\usepackage[usenames,dvipsnames]{xcolor}
\usepackage[right=3cm,left=3cm,top=4cm,bottom=3cm,headsep=0cm,footskip=0.5cm]{geometry}
\usepackage{amsmath,tabu}
\usepackage{mathtools}
\usepackage{amssymb,latexsym,mathrsfs,amstext} 
\usepackage{array}
\newcolumntype{C}{>{$}c<{$}}
\DeclareMathOperator{\Tr}{Tr} 
\usepackage{leftidx}
\usepackage{slashed}
\usepackage{natbib}
\usepackage{anyfontsize}
\usepackage{t1enc}
\usepackage{layout} 
\usepackage{verbatim} 
\usepackage{graphicx} 
\usepackage{wrapfig} 
\usepackage{anyfontsize}
\usepackage{calligra}
\usepackage{cancel}
\DeclareGraphicsExtensions{.jpg,.eps,.jpeg,.pdf,.png,.2} 
\usepackage{float}
\newcommand{\mc}{\mathcal}

\newcommand{\dg}{\dagger}
\def\eq#1{{eq.~(\ref{#1})}}
\def\eqs#1#2{{eqs.~(\ref{#1})--(\ref{#2})}}

\def\Tr{\mbox{Tr}\,}

\def\FP{fixed point }
\def\FPs{fixed points }

\def\ie{{\it i.e.}}

\definecolor{oucrimsonred}{rgb}{0.6, 0.0, 0.0}
\definecolor{persianblue}{rgb}{0.11, 0.22, 0.73}
\definecolor{forestgreen}{rgb}{0.13,0.35,0.13}
 \hypersetup{colorlinks, citecolor=oucrimsonred, linkcolor=persianblue, urlcolor=oucrimsonred}
\newcommand{\be}{\begin{equation}}
\newcommand{\ee}{\end{equation}}
\newcommand{\bea}{\begin{eqnarray}}
\newcommand{\eea}{\end{eqnarray}}
\newcommand{\nn}{\nonumber}

\begin{document}
\title[]{ In search of a UV completion of the Standard Model \\[-0.6em]
{\small 378,000 models that don't work}}
\date{\today}
\author{D.\ Barducci$^{\dag\ddag}$}
\author{M.\ Fabbrichesi$^{\ddag}$}
\author{C.\ M.\ Nieto$^{\dag\ddag\ast}$}
\author{R.\ Percacci$^{\dag\ddag}$}
\author{V.\ Skrinjar$^{\dag}$}
\affiliation{$^{\dag}$Scuola Internazionale di Studi Superiori Avanzati (SISSA), via Bonomea 265, 34136 Trieste, Italy}
\affiliation{$^{\ddag}$INFN, Sezione di Trieste, via Valerio 2, 34127 Trieste, Italy}
\affiliation{$^{\ast}$ICTP,  Strada Costiera 11, 34151 Trieste, Italy}

\begin{abstract}
\noindent  \vskip 0.8cm
Asymptotically safe extensions of the Standard Model have been 
searched for by adding vector-like fermions charged under the Standard Model gauge group and having Yukawa-like interactions with new scalar fields. 
Here we study the corresponding renormalization group $\beta$-functions to next and next-to-next to leading order in the perturbative expansion, varying the number of extra fermions and the 
representations they carry. 
We test the fixed points of the $\beta$-functions against various criteria of perturbativity to single out those that are potentially viable.
We show that all the candidate ultraviolet fixed points 
are unphysical for these models: 
either they are unstable under radiative corrections, or they cannot be matched to the Standard Model at low energies.
\end{abstract}
\maketitle

{ \setstretch{1.0} \small \tableofcontents}

\section{Introduction}
\label{sec:intro}

In the early days of quantum field theory, renormalizability was used as a criterion
to select physically viable models.
It was later understood that effective field theories  can be useful and predictive in their domain of validity even if not renormalizable.
Current particle physics is largely based on this paradigm.
It leaves an enormous freedom.
One would like to have a more restrictive framework to guide
the search for physics beyond the Standard Model (BSM).
Non-perturbative renormalizability,  also known as {\it asymptotic safety} (AS), provides such a framework.
A quantum field theory is AS if all its  couplings, running along  the renormalization group (RG) flow, reach a fixed point  in the ultraviolet (UV) limit \cite{Wilson:1971bg,Weinberg:1980gg}. 
The \FP can be interacting or free (Gaussian). In the latter case, AS reduces to asymptotic freedom (AF). 
In both cases, the theory is well behaved and predictive at all energies.
UV completeness is by itself a rather abstract notion, being untestable in practice. The real bonus of AS is that when a suitable \FP exists, typically there are only a finite number of relevant
directions that can be used to reach it in the UV.
This greatly restricts the infrared physics.

While AF theories have been studied in great detail and for a long time, work on AS models for particle physics has only begun quite recently.
For some early references based on the use of the functional
renormalization group see \cite{Gies:2009hq,Gies:2009sv,Fabbrichesi:2010xy,Bazzocchi:2011vr,Fabbrichesi:2011bx,Gies:2013pma}.
A breakthrough came with the work of Litim and Sannino,
who constructed gauge-Yukawa systems admitting interacting \FPs 
that are under perturbative control \cite{ASI}.
In these models the \FP arises from a cancellation between
one- and two-loop terms in the $\beta$-functions.
The crucial ingredient is the Veneziano limit,
providing the small expansion parameter
\be
\epsilon=\frac{N_f}{N_c}-\frac{11}{2}\ ,
\ee
where $N_c$ and $N_f$ are the numbers of colors and flavors
respectively.
It is reasonable to expect that there may exist AS models
also for finite values of $\epsilon$.
General conditions for the existence of such \FPs
have been discussed in \cite{TheoremsAS,ASI}.
Applications of these ideas to BSM physics have appeared
\cite{Litim:2015iea,Bajc:2016efj,Abel:2017ujy,Abel:2017rwl,Litim,Mann:2017wzh,Pelaggi:2017abg,Pelaggi:2017wzr,Ipek:2018sai}.

The Standard Model (SM) by itself is not AS because of the Landau pole in the $U(1)$ gauge coupling \cite{GellMann,Gockeler} and the uncertain fate of the Higgs quartic interaction \cite{Degrassi}.
The Landau pole can only be avoided by assuming that the
gauge coupling is identically zero at all energies. 
This is known as the triviality problem.

Can the SM be turned into an AS  theory by extending its  matter content? 
The simplest (and most studied) extensions consist of multiple generations of vector-like fermions carrying diverse representations of the SM gauge group. The choice of vector-like fermions is motivated by their not giving rise to gauge anomalies and their masses being technically natural.
The authors of \cite{Litim} have studied the $\beta$-functions to two-loop order  in  the simplified case of $SU(3)\times SU(2)$ 
gauge interactions and a Yukawa-like interaction among the vector-like fermions. They find several UV fixed points, which they match to the low-energy SM in a number of benchmark cases.
In a parallel development, the authors of \cite{Mann:2017wzh,Pelaggi:2017abg} studied AS for the full SM gauge group, again extended by vector-like fermions, 
by means of a resummation of the perturbative series of the $\beta$-functions.  
They find several UV fixed points, which however cannot be matched to the low-energy SM in a consistent manner~\cite{Pelaggi:2017abg}.

To move forward in this program, 
we report  our results for a large class of models based 
on the SM matter content and with
$SU_c(3)\times SU_L(2)\times U_Y(1)$ gauge interactions,
but retaining
only the top Yukawa coupling and the Higgs quartic self-interaction;
in addition, the models contain fermions 
that are coupled to the SM gauge fields via vector currents 
and have Yukawa interactions with a new set of scalar fields.
The restriction to the top Yukawa makes the form of the
$\beta$-functions more manageable---and is in line with earlier investigations.
The models differ in the number of copies of the vector-like fermions
and in the representation of the gauge groups that they carry.

In contrast to \cite{Mann:2017wzh,Pelaggi:2017abg} we do not use resummed $\beta$-functions.
Instead, we compare the results of the two-loop (NLO) gauge $\beta$-functions
considered so far in the literature
with the three loop results (NNLO).
As explained in section \ref{sec:1},
the $\beta$-functions for the Yukawa and scalar couplings
are retained always at one- and two-loops less than the gauge couplings, respectively.
By comparing the results at these two different approximation schemes,
we are able to assess quantitatively the impact of
radiative corrections and therefore to decide whether a given fixed point is within the
perturbative domain or not. This selection is supported by the use of other tests of
perturbativity that the fixed points must satisfy, as discussed in sections \ref{sec:tests} and \ref{sec:CC}.

The core of our work consists of a systematic search of reliable \FPs
in a large grid parameterized by the value of $N_f$ (the number of vector-like fermions)
and their $SU(3)_c\times SU(2)_L\times U(1)_Y$ quantum numbers
(this grid  is defined precisely in section \ref{sec:mod}).
We first find all the zeroes of the $\beta$-functions for each model in the grid.
We then test each \FP thus found against two conditions:
\begin{itemize}
\item   The \FP must occur in a region in which the perturbative expansion is reliable. 
At the very least, this implies that it must be possible to reasonably trace 
 its value at  the NNLO back to that of the NLO.
We  see {\it a posteriori} that this can be done only when the values of the couplings
and of the scaling exponents (the eigenvalues of the linearized expansion around the fixed point) are sufficiently small and the \FP satisfies all the criteria introduced in section \ref{sec:1}.
\item The \FP can be connected to the SM at low energy.
In general this would require a delicate numerical analysis of the trajectories 
emanating from it. However, we find that a rough necessary condition is sufficient for
our purposes: the \FP must not have any coupling that is zero and irrelevant,
because such couplings must be identically zero at all scales to avoid Landau poles.
\end{itemize}

As we shall see, these two requirements taken together, while quite reasonable, are very restrictive.
As a matter of fact, we are not able to identify {\it any} choice for the group representations and number of generations of the vector-like fermions that would make the extension of the SM reliably AS. 
This does not mean that such an extension does not exist: 
it only means that if such an AS extension of the SM exists, 
it must either be different from those that we have considered,
or else it must have a \FP that lies outside the reach
of perturbation theory.

\section{Methods}
\label{sec:1}

In this section we describe the general procedure that 
we follow in the rest of the paper. 
This  allows us to motivate better the requirements (introduced in section \ref{sec:intro} and further elaborated here) that we impose on the \FPs in order for them to be considered as physical.
We recommend \cite{Hollowood} as a general reference on RG flows.

\subsection{The fixed points of the $\beta$-functions}
\label{sec:FP}

Consider a theory with generic (gauge, fermion or scalar) fields   
and (generally dimensionful) couplings $\bar g_{i}$ of the interactions among them.
In the study of the RG it is customary to use
dimensionless couplings $g_i$, 
related to the dimensionful couplings by 
$g_{i}=\mu^{-d_{i}}\bar{g_{i}}$,
$d_{i}$ being the mass dimension of $\bar{g_{i}}$. 
The renormalization of the theory is completely characterized by its 
$\beta$-functions
\be
 \beta_i (g_j)\equiv \mu \frac{d g_i}{d\mu}\, ,
 \ee
 where $\mu$ is the sliding scale of the quantum theory.
A \FP of this theory, denoted $g^{*}_{j}$, is defined by the vanishing of the  $\beta$-functions of all couplings:
\be
\beta_{i} (g_j^*)=0\,.
\ee

When the  couplings $g_{j}$ assume the values $g^{*}_{j}$, the renormalization of the quantum theory stops. 
In general, a given \FP can be reached either in the UV or in the
IR limit, depending on the direction of the approaching trajectory.
Notice that the familiar distinction between UV and IR \FPs
is only meaningful when there is a single coupling in the theory.

The $\beta$-function of a single coupling is
independent of the gauge choice in dimensional regularization. 
It is regularization scheme-independent up to NLO. If there are several  couplings running together, their $\beta$-functions depend on the scheme already at the NLO~\cite{McKeon:2017mjq}. There is therefore a degree of ambiguity in the position of the \FPs we are going to discuss because their position could be moved by changing the scheme. We assume that these changes are small if the \FP is found within the perturbative regime. One should however bear in mind this problem of scheme dependence in all the discussions to follow.

In general, there are no conditions on the values of the \FP $g^{*}_{i}$ and they could take any value. However, when we work in perturbation theory, we have to remain within its range of validity. Therefore, we demand all the  coupling at the \FP $g^{*}_{i}$ to be sufficiently small.
In practice this means that going to the next order of the expansion
should not change appreciably the position of the fixed point as well as its other properties.
We will see that  this implies that the numerical value of the \FPs must satisfy the condition
\be
\alpha^{*}_{i} \equiv   \left(\frac{g^{*}_{i}}{4\pi}\right)^{2}  <1 \label{def} \, ,
\ee
in addition to being positive.
Notice that in \eq{def}, and in what follows, the definition of the coupling $\alpha$  follows a convention widely adopted in the AS literature which however differs from the usual one by an additional factor $4 \pi$
in the denominator.

The condition in \eq{def} would suffice to keep the perturbative expansion within its limits of validity if the coefficients in the perturbative expansions were of the same order and not too large. If they are not, the condition in \eq{def} should be strengthened and only smaller values allowed to prevent terms of higher order to be more important than those at the lower order. 

As we shall see, this is the case for many of the \FPs we discuss. As a matter of fact, many of the \FPs discussed in the literature are due to a cancellation between the first two orders in the perturbative expansion of the $\beta$-functions.  This is acceptable only if  the  higher terms in the perturbative expansion are then more and more suppressed. This is the main motivation for going to three-loop order in the gauge $\beta$-functions.

\subsection{Linearized flow}
\label{sec:LF}

Once we have a candidate fixed point, we can study the flow in its immediate neighborhood. 
We move away from the \FP and study what happens when we shift
the couplings by a small amount $y_{i}\equiv g_{i}-g^{*}_{i}$. To this end, we linearize the $\beta$-functions as
\begin{equation}
    \frac{dy_{i}}{dt}=M_{ij}y_{j}\, ,
    \label{Stability}
\end{equation}
 where $M_{ij}\equiv \partial\beta_{i}/\partial g_{j}$
is referred to as the {\it stability matrix}. Next, we diagonalize the linear system by going to the variables $z_{i} =(S^{-1})_{ij}y_{j}$,
defined by the equation
\begin{equation}
    (S^{-1})_{ij}M_{jl}S_{ln}=\delta_{in}\vartheta_{n} \, ,
\end{equation}
so that the $\beta$-functions and their solutions are in the simplified form  
\begin{equation}
    \frac{dz_{i}}{dt}=\vartheta_{i}z_{i}\, \quad \mbox{and} \quad \ z_{i}(t)=c_{i}\, e^{\vartheta_{i}t}=c_{i}\left( \frac{\mu}{\mu_{0}} \right)^{\vartheta_{i}}.
\end{equation}

From the expression of $z_{i}$ as functions of $\mu$, we see that there are different situations depending on the sign of $\vartheta_{i}$:
\begin{itemize} 

\item For $\vartheta_{i}>0$, as we increase $\mu$ we move away from the \FP and $z_{i}$ increases without control; the direction $z_{i}$ is said to be {\it irrelevant}.

\item If $\vartheta_{i}<0$, as we increase $\mu$ we approach the fixed point; 
 the direction $z_{i}$ is called a {\it relevant} direction.
 
\item If $\vartheta_{i}=0$, we do not know the fate of $z_{i}$ and we have to go beyond the linear order as explained below;
the direction $z_{i}$ is called {\it marginal} in this case.
\end{itemize}
 
\begin{figure}[t]
\centering
 \includegraphics[scale=1.2]{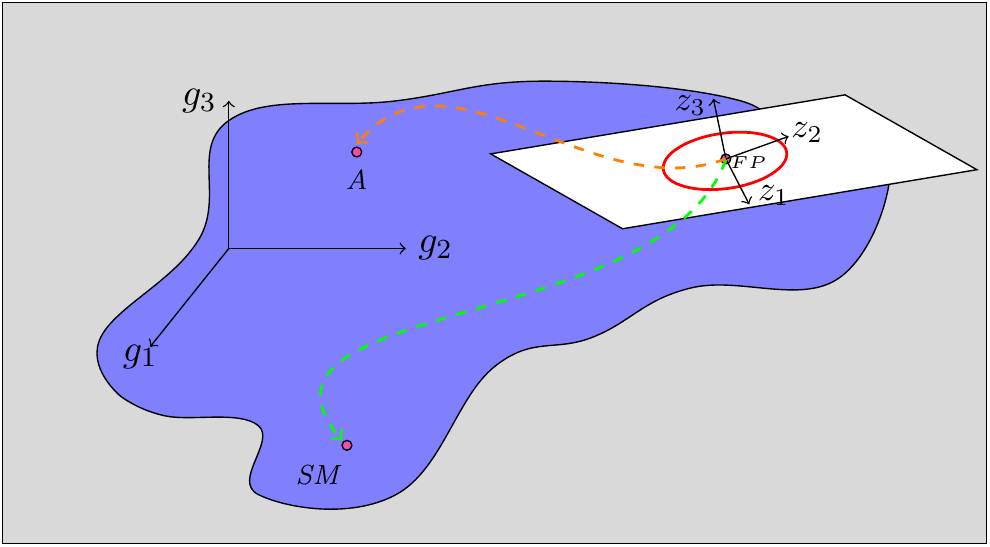}
\caption{ \small \setstretch{1.0} Theory space of couplings $g_{i}$ where only $3$ axes are shown for simplicity. For a given \FP we show the UV 
safe surface ({\color{blue} blue} region), the approximated UV critical surface around the \FP (white plane), the new set of coordinates $z_{i}$, 
a small region of possible initial points for the flow ({\color{red} red} circle) and two UV safe trajectories ending at a given matching scale 
$\mc{M}$ ({\color{green} green} and {\color{orange} orange} dashed lines, the former going to the SM, the latter going to a different IR physics A).}
\label{fig:Phasespace}
\end{figure}

The notion of relevance/irrelevance is independent
of the direction of the flow and of the choice of basis. AS theories correspond to trajectories
lying on the surface whose tangent space at the \FP
is spanned by the relevant directions. This tangent space is shown in Figure \ref{fig:Phasespace} as a white plane. In the same figure we depict the full UV critical surface in blue.

The eigenvalues $\vartheta_{i}$ have the property of being universal quantities---meaning that they are invariant under a general transformation in the space of couplings. In perturbation theory, they cannot take any arbitrary value, there are restrictions on their size. We know that in general the $\beta$-function for $g_{i}$ has the form
\begin{equation}
    \beta_{i}=-d_{i}g_{i}+\beta^{q}_{i}(g_{j}),
    \label{Generalbeta}
\end{equation}
where $\beta^{q}_{i}$ encodes the pure quantum contributions to the $\beta$-functions. Therefore, the stability matrix is given by
\begin{equation}
    M_{ij}=-d_{i}\delta_{ij}+\frac{\partial\beta^{q}_{i}}{\partial g_{j}}
\end{equation}
which is equal to the classical scaling exponent plus quantum corrections. Then, the quantity $-\vartheta_{i}$ represents the full  scaling dimension of the coupling $\bar{g_{i}}$. If we want to remain in perturbation theory, we should demand 
that the scaling dimension be small. 
In the cases we  consider, where all the couplings have $d_{i}=0$, this means that 
\begin{equation}
|\vartheta_{i}|<O(1)\, .
\label{thetabound}
\end{equation}
There is a degree of arbitrariness about where exactly one should set this bound. In our study, we look at the  scaling dimensions for the models under examination and set the bound in the first gap in the distribution of their $O(1)$ values.

\subsection{Marginal couplings}
\label{sec:marg}

If one of the eigenvalues is equal to zero, the linear approximation does not give us information about the RG behaviour in the direction associated to it. Then we have to go further in the expansion
 \cite{Codello:2016muj}.
At second order in the couplings $y_{i}$, the $\beta$-functions take the form
\begin{equation}
   \frac{dy_{i}}{dt}=M_{ij}y_{j}+P_{ijk}y_{j}y_{k}\, , \quad \mbox{where} \quad  P_{ijk}=\frac{\partial^{2}\beta_{i}}{\partial g_{j}\partial g_{k}}.
    \label{QForm}
\end{equation}
The structure of these quadratic flows is quite complicated 
to describe in full generality, with the fate of a specific
trajectory depending strongly on the position of the initial
point in the neighbourhood of the fixed point.

However, marginal couplings do not generally occur for
a fully interacting fixed point: they can always be identified with some
coupling that is itself zero at the fixed point.
We show in Appendix \ref{sec:appmarg} that the structure of the $\beta$-functions is such that
the flow of the marginal couplings near the \FP is of the form
\begin{equation}
\frac{dy_{i}}{dt}=P_{iii}y_i^2\ ,
\end{equation}
(no summation implied).
Our flows will be written always in terms
of the $\alpha_i$, which are bound to be positive.
Therefore, marginal directions $y_i=\alpha_i$
with $P_{iii}<0$ are UV attractive
and are called {\it marginally relevant}
(a well-known example being the QCD gauge coupling)
while those with $P_{iii}>0$ are UV repulsive
and are called {\it marginally irrelevant}.
Altogether,  the UV critical surface is thus spanned by the relevant and marginally relevant directions.

\subsection{Standard Model matching}
\label{sec:matching}

Once we have an understanding of the \FP structure---and the conditions on the couplings $\alpha_{i}$ and the scaling exponents $\vartheta_{i}$ are satisfied---there remains to find the trajectory connecting a given \FP to the SM value of the coupling at some IR scale. 
This is accomplished in the following manner.

First, all the SM couplings are run to a common RG scale, which we take to be 1.83 TeV, using the SM $\beta$-functions, to the values
$$
\alpha_1=0.000795\ ,\ \
\alpha_2=0.00257\ ,\ \
\alpha_3=0.00673\ ,\ \
\alpha_t=0.00478\ .
$$
This defines the target for the flow to the IR from the UV fixed point.
Then, the RG flow is started from a point belonging to the UV critical surface, infinitesimally close to the \FP (red circle in Figure \ref{fig:Phasespace}).
This guarantees that, to high precision, the flow towards the UV
ends at the fixed point.
The system  is then allowed to flow by means of the full $\beta$-functions of the theory towards the IR.
The initial point of the flow
is varied until, ideally, the trajectory hits exactly the target SM values.

The scale at which one starts the flow, $\mu_0$, is not known a priori.
If one reaches the target values of the couplings after
some RG time $t=\log(\mu/\mu_0)<0$ (in our conventions $t$ increases towards the UV), 
the corresponding scale $\mu$ is identified with 1.83  TeV 
and the starting scale is identified as $\mu_0=\mu e^{-t}$.

For most of the models that we consider, this laborious procedure is not necessary.
For all their \FPs that can be regarded as being in the perturbative domain,
the hypercharge is zero at the \FP
and is also a marginally irrelevant coupling.
This means that in order to reach the \FP in the UV limit,
the hypercharge must be zero at all energies.
All other trajectories have a Landau pole.
These models are thus excluded by a version of the triviality problem.

\subsection{Approximation schemes}
\label{sec:schemes}

The perturbative $\beta$-functions of the SM and its extensions have a natural hierarchy 
originating from the  Weyl consistency conditions \cite{Osborn,Cardy,Jack,OsbornII,Antipin}: 
\be
\frac{\partial \beta^j}{\partial g_i} =\frac{\partial \beta^i}{\partial g_j} \, . \label{WC}
\ee
A consistent solution of \eq{WC} relates different orders in the perturbative expansion and indicates that 
the gauge couplings must have the highest order in the loop expansion, while the Yukawa coupling must be computed at one order less and the quartic interaction one further order less. This leaves us in practice with two approximations for the running of the couplings: 
\begin{itemize}
\item \underline{the {\tt 210} approximation scheme}, in which the gauge couplings are renormalized at the two-loop order (NLO), the Yukawa coupling only at one-loop order (LO) and the quartic interaction is not renormalized; and  
\item \underline{the {\tt 321} approximation scheme},  in which the gauge couplings are renormalized at the three-loop order (NNLO), the Yukawa coupling at two-loop order (NLO) and the quartic interaction at one-loop order (LO).
\end{itemize}
By comparing the two approximations it is possible to test the stability of the \FP against radiative corrections and the overall reliability of the perturbative computation. 

Other approximation schemes are also possible, for example retaining all $\beta$-functions at the same order or keeping only the gauge $\beta$-functions one order higher than the others. These different choices do not satisfy \eq{WC}. 
They are analysed in \cite{Bond:2017tbw} where they respective merits (and shortcomings)  are discussed.

\subsubsection{ Perturbative $\beta$-functions: A digest of the  literature}

The perturbative study of the $\beta$-functions of the SM, together with some of its possible extensions, has been a collective endeavor covering many years. We collect here  the main stepping stones in this ongoing computation. 

The one-loop (LO) $\beta$-function for  a non-abelian gauge group was computed in the classic papers 
\cite{Gross:1973id} and \cite{Politzer:1973fx} where AF was discovered. The LO $\beta$-function for the Yukawa coupling was presented in \cite{Cheng:1973nv} and that for the quartic Higgs interaction in \cite{Gross:1973ju}.
The two-loop (NLO) $\beta$-functions for the  gauge groups have been calculated
in \cite{Caswell:1974gg,Tarasov:1976ef,Jones:1981we,MachacekI}, those for the  Yukawa couplings in \cite{Fischler:1982du,MachacekII,Jack:1983sk} and that for the quartic Higgs interaction in \cite{MachacekIII,Jack:1983sk,Ford:1992pn}. The case of
 the SM has been discussed in \cite{Arason:1991ic}.
Mistakes in some of these results were corrected in \cite{Luo,LuoII} where they were also generalized to arbitrary representations of non-simple groups. 
The three-loop (NNLO) $\beta$-functions of a gauge
theory with simple groups were given partially in \cite{Curtright:1979mg}, then in \cite{Pickering}. 
The full NNLO $\beta$-functions for the SM were presented in \cite{Mihaila:2012fm}
and those for generic representations of non-simple gauge groups
in \cite{Mihaila}.
In this last paper, some contributions from the Yukawa and quartic Higgs interactions were not included. 
For these terms we have used currently unpublished results of L.~Mihaila \cite{Mihaila2}.
The NNLO $\beta$-functions for the Yukawa  and  quartic Higgs couplings were  partially computed in \cite{Chetyrkin:2012rz} and fully in \cite{Bednyakov:2012en,Bednyakov:2013eba}. We will not need them here.

\subsection{Another test of perturbativity}
\label{sec:tests}

Besides the smallness of the couplings themselves,
there is another  simple test that we  use to
assess whether a \FP is in the perturbative domain.

Let us write the $\beta$-functions of the gauge couplings $\alpha_i$ 
in the schematic form
\be
\beta_i=\Big(A^{(i)}+B^{(i)}_{r}\alpha_{r}+C^{(i)}_{rs}\alpha_{r}\alpha_{s} \Big)\alpha_i^{2},
\label{Loops}
\ee
where $A$, $B$ and $C$ are the one-, two- and three-loops coefficients.
At a \FP we can split each beta function in the following way
\be
0=\beta_i=A^{(i)}_*+B^{(i)}_*+C^{(i)}_*\ ,
\label{LoopsFP}
\ee
where  $A^{(i)}_*=A^{(i)}\alpha_{i*}^2$, 
$B^{(i)}_{*}=B^{(i)}_r\alpha_{r*}\alpha_{i*}^2$ and
$C^{(i)}_{*}=B^{(i)}_{rs}\alpha_{r*}\alpha_{s*}\alpha_{i*}^2$.
When we insert the values of the \FP calculated in the {\tt 321}
approximation scheme, we expect the three contributions
to be ordered as
$|C^{(i)}_*|<|B^{(i)}_*|<|A^{(i)}_*|$,
or equivalently
\be
\rho_i<\sigma_i<1
\ ,\quad
\mathrm{where}
\quad 
\rho_i=|C^{(i)}_*/A^{(i)}_*|
\quad\mathrm{and}\quad
\sigma_i=|B^{(i)}_*/A^{(i)}_*|\ .
\label{LoopsFPbound}
\ee
Note that in this case $\rho+\sigma=1$.
In priciple it might also happen that the sum of the one-
and three-loop terms cancel the two-loop term,
{\it i.e.} $\sigma-\rho=1$. We shall see that this does not happen.
Alternatively, if we insert in (\ref{Loops}) the \FP values
of the {\tt 210} approximation scheme, $\beta_i$ will not be zero.
The first two terms in (\ref{LoopsFP}) will cancel, but we still expect
$\rho_i<1$.
In the following, when we report results in the {\tt 210} approximation scheme,
we  give the values of $\rho_i$ defined at the {\tt 210} \FP
and when we report results in the {\tt 321} approximation scheme,
we  give the values of $\rho_i$ and $\sigma_i$ defined at the {\tt 321} approximation scheme fixed point.

\subsection{Testing \FPs with central charges}
\label{sec:CC}

At a \FP the theory is a conformal field theory (CFT). As explained in appendix \ref{sec:CFT}, one  can estimate the size of the relative changes of the central charges of the CFT to decide whether a \FP is within the domain of perturbation theory. These relative changes are obtained in terms of the 
function  $a=a_{free}+a_q$ ($a_q$ refers to the contribution of quantum corrections) and of the $c$-function as
\be
\delta a \equiv \frac{a-a_{free}}{a_{free}}=\frac{a_q}{a_{free}}	\quad \mbox{and} \quad
\delta c \equiv \frac{c-c_{free}}{c_{free}}=\frac{c_q}{c_{free}}\, .
\ee
If $\delta a$ or $\delta c$ become smaller than $-1$ the \FP is unphysical because it cannot correspond to a CFT  (since $c>0$ and $a>0$ are guaranteed for CFT). 
A \FP for which $\delta c$ or $\delta a$ is of order $1$ should be  discarded as well since quantum corrections are then comparable in size to the free-theory contribution.  

The central charges in the {\tt 210} approximation scheme can be easily computed by embedding the models in the general gauge-Yukawa Lagrangian of \cite{Dondi:2017civ}.
Computation in the {\tt 321} approximation scheme is significantly more complicated due to a major increase in complexity of the Zamolodchikov metric. 
We do not pursue the {\tt 321} computation for that reason and 
also because the results in the {\tt 210} approximation scheme 
are enough to confirm that our other perturbativity  criteria
are compatible with the CFT tests.

\subsection{Procedure summary}
\label{sec:procsum}
 
Given a model, we first look for all the \FPs of the $\beta$-functions.
Since the $\beta$-functions are given in the form of
a Taylor expansion, they will have several zeroes
that are mere artifacts of the expansion,
and we have to select those that have a chance of being physical.
The criteria we apply are: stability under radiative corrections
and matching to the SM at low energy.

We begin by analyzing the \FPs of the {\tt 210} approximation scheme.
In the first step, we retain only those \FPs
that can be reasonably assumed to be within the perturbative regime, that is, those for which
the couplings and the scaling exponents satisfy
the bounds in \eq{def} and \eq{thetabound}.
We use the criteria discussed 
in sections \ref{sec:tests} and \ref{sec:CC}
to confirm that these bounds
are indeed reasonable indicators of radiative stability.

We then compare with the results of the same analysis
in the {\tt 321} approximation scheme. 
We retain only those \FPs that can be reasonably identified
in both approximations. Their number is quite small.
We find that the identification is only possible if the couplings and scaling exponents are sufficiently small.

Finally, for the \FPs that are radiatively stable in the sense
just described, we look for the possibility of matching to the SM at low energy. 
If all these conditions are satisfied, we have a \FP that can be considered as physical. Otherwise, the \FP should be rejected and deemed unphysical.

\section{The fate of the Standard Model couplings}
\label{sec:2}

The running of the SM couplings, when extended to high energies,
presents two important features: partial gauge coupling unification and a Landau pole in the  abelian gauge coupling. Since this singularity appears beyond the Planck scale, where gravitational effects are important, it might well happen that there will be no divergence and that all couplings are well-behaved once we consider a full theory of gravity and matter. Nevertheless, it is interesting to investigate whether such infinities could be avoided within the matter sector. This study will nicely illustrate our procedure by means of the familiar case of the SM.

\begin{figure}[t]
 \centering
\includegraphics[scale=0.7]{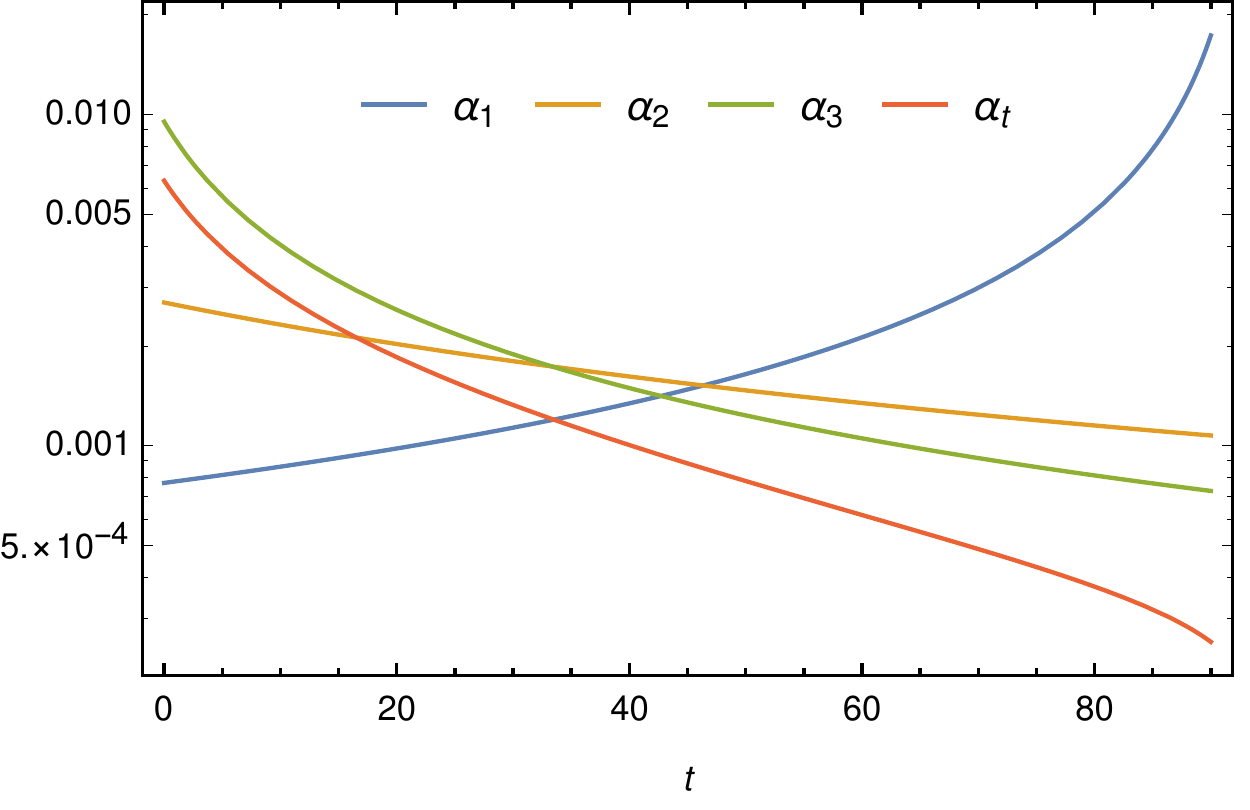}
 \caption{\small \setstretch{1.0} Running of the gauge couplings $\alpha_{i}$ and Yukawa $\alpha_{t}$  for the SM in the {\tt 321} approximation scheme.  On the horizontal axis $t=\ln \left( \mu /M_{Z} \right)$. Just above $t\simeq 40$ the three gauge couplings come close together. At larger values of $t$, $\alpha_1$ begins its ascent towards the Landau pole. }
\label{SMrunning}
\end{figure}

Throughout this paper,
we shall consider a simplified version of the SM 
where only the top-Yukawa coupling $y_{t}$ is retained.
The remaining Yukawa couplings are set to zero.
For simplicity we will keep calling this the SM.
However, we stress that the degrees of freedom that enter the flow 
are not only those of the top quark but 
the full SM matter content
(\ie, the number of fermions that enters into the 1-loop coefficient of the gauge $\beta$-functions counts all the quarks and leptons).

The first question is whether the $\beta$-functions of the SM have fixed points.
This does not happen with the LO beta functions.
In Figure~\ref{SMrunning} we show the running of the couplings toward the 
(quasi)-unification point and the 
beginning of the ascent of the coupling $\alpha_1$
toward the Landau pole.

\subsection{The {\tt 210} approximation scheme}
 
We then consider the beta functions in the {\tt 210} approximation scheme (NLO in the gauge couplings),
which are given by
\bea
    \beta_{1}^{\mbox{\tiny NLO}} &=&\alpha_{1}^{2}\left(\frac{41}{3}+\frac{199}{9}\alpha_{1}+9\alpha_{2}+\frac{88}{3}\alpha_{3}-\frac{17}{3}\alpha_{t} \right), \nn \\
    \beta_{2}^{\mbox{\tiny NLO}} &=&\alpha_{2}^{2}\left(-\frac{19}{3}+3\alpha_{1}+\frac{35}{3}\alpha_{2}+24\, \alpha_{3}-3\,\alpha_{t} \right), \nn \\
    \beta_{3}^{\mbox{\tiny NLO}} &=&\alpha_{3}^{2}\left( -14+\frac{11}{3}\alpha_{1}+9\, \alpha_{2}-52\, \alpha_{3}-4\, \alpha_{t} \right), \nn \\
    \beta_{t}^{\mbox{\tiny LO}} &=&\alpha_{t}\left( -\frac{17}{6}\alpha_{1}-\frac{9}{2}\alpha_{2}-16\, \alpha_{3}+9\, \alpha_{t} \right)  ,
    \label{sm210}
\eea
where, following the convention (\ref{def}), we use the variables
\be
\alpha_{i}=\frac{g_{i}^{2}}{(4\pi)^{2}}\ \mathrm{for}\ i=1,2,3,\quad \mbox{and} \quad
\alpha_{t}=\frac{y_{t}^{2}}{(4\pi)^{2}}\, .  
\label{couplings}
\ee
The set of $\beta$-functions in \eq{sm210} admits several zeroes.
They are given by the last column of Table \ref{tab:grand}
in Appendix \ref{sec:appbeta}.
However, only two of them (solutions $P_{16}$ and $P_{17}$)
have all $\alpha_i$ positive. 
Their properties are summarized in Table \ref{tab:SM210}.

\begin{table}[H]
    \centering
{\scriptsize  \setstretch{1.5}
\begin{tabular}{|C|C|C|C|C|C|C|C|C|}\hline
\rowcolor{Yellow}
 \hspace*{0.8cm}  & \alpha^{*}_{1} & \alpha^{*}_{2} & \alpha^{*}_{3} & \alpha^{*}_{t} &  \vartheta_{1} & \vartheta_{2} & \vartheta_{3} & \vartheta_{4} \\ \hline
\cellcolor{Yellow} FP_1 & 0 & 0.543 & 0 & 0 & 3.44 & -2.44 & 0 &  0 \\  
\cellcolor{Yellow} FP_2 & 0 & 0.623 & 0 & 0.311 & 5.21 & 2.21 & 0 & 0\\  
\hline
\end{tabular}
}
\caption{\small  \setstretch{1.0} Fixed points and their scaling exponents
for the SM in the {\tt 210} approximation scheme.}
\label{tab:SM210}
\end{table}

Although less than 1, the values for the couplings constants $\alpha_i^*$ are  quite sizeable and we may suspect that they may lie outside the perturbative domain. This suspicion is substantiated by looking at the linearized flow. 
Considering that these exponents are classically zero,
we see that the quantum correction are quite large.
The fate of the marginal directions ($z_{3}$ and $z_{4}$) is determined by looking at the quadratic approximation to the flow,
as discussed in section \ref{sec:marg} and in appendix \ref{sec:appmarg}.
We find that the third direction is marginally irrelevant while the last one is marginally relevant.

Even if we had decided to ignore the breaking of the perturbative regime and insisted on looking for trajectories connecting one of the \FPs to the IR regime, the requirement of 
lying on the UV critical surface would have implied that there is always a coupling that vanishes at all scales. Namely, given that $\alpha^{*}_{1}=0$, and that the $\beta$-function for $\alpha_{1}$ is proportional to a power of $\alpha_{1}$ itself, this coupling cannot run at all. In other words, the coupling $\alpha_{1}$ is frozen at zero at all scales and the $U(1)$ gauge interaction is trivial. 
Clearly there are no physical \FP within the SM in the {\tt 210} expansion:
the problem of the Landau pole is still present even when the gauge couplings
are taken at NLO.

\subsection{The {\tt 321} approximation scheme}

To check the perturbative stability of the two \FPs of the previous section, 
we now study the $\beta$-functions to the next order. 
In the {\tt 321} approximation scheme (NNLO in the gauge couplings), 
the $\beta$-functions take the form \cite{Antipin}
\bea
\beta_{1}^{\mbox{\tiny NNLO}}&=&\beta_{1}^{\mbox{\tiny NLO}}+
\alpha_{1}^{2}\left[
-\frac{388613}{2592}\alpha_{1}^{2}+\frac{205}{48}\alpha_{1}\alpha_{2}+\frac{1315}{32}\alpha_{2}^{2}-\frac{274}{27}\alpha_{1}\alpha_{3} 
-2\, \alpha_{2}\alpha_{3}+198\, \alpha_{3}^{2}
\right.\nonumber \\ 
& & 
-\left( \frac{2827}{144}\alpha_{1}+\frac{785}{16}\alpha_{2}+\frac{58}{3}\alpha_{3} \right)\alpha_{t}+\frac{315}{8}\alpha_{t}^{2} 
\left.\vphantom{\frac{1}{2}} +\frac{3}{2}\Big( \alpha_{1}+\alpha_{2}-\alpha_{\lambda} \Big)\alpha_{\lambda}   \right], \nn \\
\beta_{2}^{\mbox{\tiny NNLO}} &=&\beta_{2}^{\mbox{\tiny NLO}}+ 
\alpha_{2}^{2}\left[ 
-\frac{5597}{288}\alpha_{1}^{2}+\frac{291}{16}\alpha_{1}\alpha_{2}+\frac{324953}{864}\alpha_{2}^{2}-\frac{2}{3}\alpha_{1}\alpha_{3} 
+78\, \alpha_{2}\alpha_{3}
+162\,  \alpha_{3}^{2}
\right.\nonumber \\
    & &\left.\vphantom{\frac{1}{2}} 
-\left( \frac{593}{48}\alpha_{1}+\frac{729}{16}\alpha_{2}+14\alpha_{3} \right)\alpha_{t}+\frac{147}{8}\alpha_{t}^{2}+\frac{1}{2}\Big( \alpha_{1}+3\alpha_{2}-3\alpha_{\lambda} \Big)\alpha_{\lambda}   \right], \nn \\
\beta_{3}^{\mbox{\tiny NNLO}} &=& \beta_{3}^{\mbox{\tiny NLO}}+
\alpha_{3}^{2} \left[ 
-\frac{2615}{108}\alpha_{1}^{2}
+\frac{1}{4}\alpha_{1}\alpha_{2}
+\frac{109}{4}\alpha_{2}^{2}
+\frac{154}{9}\alpha_{1}\alpha_{3} 
+42\, \alpha_{2}\alpha_{3}+65\, \alpha_{3}^{2}
\right.\nonumber \\
& & \left.\vphantom{\frac{1}{2}}
-\left( \frac{101}{12}\alpha_{1}+\frac{93}{4}\alpha_{2}+80\alpha_{3} \right)\alpha_{t}+30\, \alpha_{t}^{2} \right] \, ,
%
\nn\\
\beta_{t}^{\mbox{\tiny NLO}}&= &\beta_{t}^{\mbox{\tiny LO}}+
\alpha_{t}\left[ 
+\frac{1187}{108}\alpha_{1}^{2}-\frac{3}{2}\alpha_{1}\alpha_{2}-\frac{23}{2}\alpha_{2}^{2}+\frac{38}{9}\alpha_{1}\alpha_{3}  +18\, \alpha_{2}\alpha_{3}-216\, \alpha_{3}^{2}
\right.\nonumber \\
    & & \left.\vphantom{\frac{1}{2}}
    +\left( \frac{131}{8}\alpha_{1}+\frac{225}{8}\alpha_{2}+72\alpha_{3} \right)\alpha_{t}-24\, \alpha_{t}^{2}-12\, \alpha_{t}\alpha_{\lambda}+3\, \alpha_{\lambda}^{2} \right],  \\
  \beta_{\lambda}^{\mbox{\tiny LO}} &=&12\alpha_{\lambda}^{2}-\Big(3\alpha_{1}+9\alpha_{2}\Big)\alpha_{\lambda}+\frac{9}{4}\left(\frac{1}{3}\alpha_{1}^{2}+\frac{2}{3}\alpha_{1}\alpha_{2}+\alpha_{2}^{2}\right)+12\, \alpha_{t}\alpha_{\lambda}-12\, \alpha_{t}^{2}, \nn
    \label{BetalambdaSM}
\eea
where the quartic Higgs coupling 
\be
\alpha_{\lambda}=\frac{\lambda}{(4\pi)^{2}}
\ee
is no longer unrenormalized.

Due to the higher order of the equations, there are more  fixed points than the two found in the {\tt 210} approximation scheme. They are listed in Table \ref{tab:SM321}.

\begin{table}[h]
    \centering
{\scriptsize  \setstretch{1.5}
\begin{tabular}{|C|C|C|C|C|C|C|C|C|C|C|}\hline
\rowcolor{Yellow}
 \hspace*{0.8cm}  & \alpha^{*}_{1} & \alpha^{*}_{2} & \alpha^{*}_{3} & \alpha^{*}_{t} & \alpha^{*}_{\lambda} & \vartheta_{1} & \vartheta_{2} & \vartheta_{3} & \vartheta_{4} & \vartheta_{5} \\ \hline
\cellcolor{Yellow} FP_1 & 0 & 0 & 0 & 0.297 & 0.184 & 8.32 & -2.57 & 0 & 0 & 0\\ 
\cellcolor{Yellow} FP_2 & 0 & 0.120 & 0 & 0.0695 & 0.0575 & 1.46 & 1.18 & 0.495 & 0 & 0 \\  
\cellcolor{Yellow} FP_3 & 0 & 0.124 & 0 & 0.333 & 0.230 & 8.82 & -2.52 & 1.38 & 0 & 0\\ 
\cellcolor{Yellow} FP_4 & 0.436 & 0.146 & 0 & 0.648 & 0.450 & -27.0 & 17.3 & -7.85 & 2.19 & 0\\  
\cellcolor{Yellow} FP_5 & 0.433 & 0 & 0 & 0.573 & 0.377 & -25.6 & 15.7 & -6.85 & 0 & 0\\
\hline
\end{tabular}
}
\caption{\small  \setstretch{1.0} Fixed points and their scaling exponents
for the SM in the {\tt 321} approximation scheme.}
\label{tab:SM321}
\end{table}
Consistently with the  discussion in the case of the {\tt 210} approximation scheme,  neither the couplings nor the exponents are small.
Moreover, it is not possible to recognize among the new \FPs 
those of the {\tt 210} approximation scheme: the values change dramatically, contrary to what would be expected in a well-behaved perturbative expansion. 

That there is a problem  is confirmed by looking at the criteria of perturbativity
introduced in section \ref{sec:tests}.
In the {\tt 210} approximation scheme,
for the two  \FPs of Table \ref{tab:SM210}, we have 
$B^{(2)}_*=1.87$ and $B^{(2)}_*=2.46$, respectively,
while $C^{(2)}_*=32.7$ and $C^{(2)}_*=53.9$, respectively. 
For both \FPs the ratio $\rho_2$ is of order 10,
grossly violating the bound (\ref{LoopsFPbound}).
It therefore appears that we are outside the domain where perturbation
theory can be trusted.

We conclude that the SM (at least in the simplified form considered here)
does not have a physical \FP within perturbation theory. 
In the next section, we study a family of models that represents the simplest 
extension to the SM content with the potential of generating new fixed points.

\section{Standard Model extensions}
\label{sec:mod}

In this section, we consider (minimal) extensions of the SM by adding new matter fields charged under the SM group $SU_c(3)\times SU_L(2)\times U_Y(1)$.
The gauge sector is not modified.
Following \cite{TheoremsAS,ASI,Litim,LitimII}, we take $N_{f}$ families of vector-like fermions minimally coupled to the SM. The idea is to consider a new type of Yukawa interactions among the vector-like fermions such that their contribution generate new zeros in the gauge $\beta$-functions. Accordingly, new scalar fields must be included as well. These scalars are taken to be singlets of the SM group while the fermions carry the representations $R_{3}$ under $SU_c(3)$, $R_{2}$ under $SU_L(2)$, and have hypercharge $Y$ of the gauge group $U_Y(1)$. Denoting $S_{ij}$ the matrix formed
with $N_f^2$ complex scalar fields, the Lagrangian characterizing this minimal BSM extension is
\begin{equation}
    \mathcal{L}=\mathcal{L}_{SM}+\Tr(\bar{\psi}i\slashed{D}\psi)+\Tr(\partial_{\mu}S^{\dg}\partial_{\mu}S)-y\Tr(\bar{\psi}_{L}S\psi_{R}+\bar{\psi}_{R}S^{\dg}\psi_{L}). 
    \label{Lagrangian}
\end{equation}
In \eq{Lagrangian}, $\mc{L}_{SM}$ stands for the SM lagrangian, $y$ is the BSM Yukawa coupling, which we assume to be the same for all fermions,
the trace  sums over the SM representation indices 
as well as the flavour indices, 
and we have decomposed $\psi$ as $\psi=\psi_{L}+\psi_{R}$ 
with $\psi_{R/L}=\frac{1}{2}(1\pm\gamma_{5})\psi$. 
We neglect the role of quartic self interactions of the scalars $S_{ij}$ 
as well as portal couplings of the latter to the Higgs sector. 

This extension of the SM is simple enough to allow explicit computations while giving rise to new features in the RG flow.
The vector-like fermions are a proxy for more elaborated extensions; they do not introduce gauge anomalies and do not induce a large renormalization of the Higgs mass:  they are technically natural.

\subsection{The $\beta$-functions}

Within the model defined by the Lagrangian (\ref{Lagrangian}), we look for \FPs satisfying the requirements discussed in section \ref{sec:procsum}. We start the analysis in the {\tt 210} approximation scheme  and write the $\beta$-functions of the system (\ref{Lagrangian}) in terms of the quantities  in \eq{couplings} augmented by the new coupling
$\alpha_{y}=\frac{y^{2}}{(4\pi)^{2}}$.

In the following, as in section \ref{sec:2}, we keep only the top-Yukawa coupling. The $\beta$-functions will depend on the dimensions of the fermion representations $d$, their Casimir invariants $C$
and Dynkin indices $S$, which are defined in general as
{\small
\begin{align}
    d_{R_{2}}&=2\ell+1, & d_{R_{3}}&=\frac{1}{2}(p+1)(q+1)(p+q+2),\nn \\
C^{(2)}_{F}&=C_{R_{2}}=\ell(\ell+1), & C^{(3)}_{F}&=C_{R_{3}}=p+q+\frac{1}{3}(p^{2}+q^{2}+pq), \nn \\
S^{(2)}_{F}&=S_{R_{2}}=\frac{d_{R_{2}}C_{R_{2}}}{3}, & S^{(3)}_{F}&=S_{R_{3}}=\frac{d_{R_{3}}C_{R_{3}}}{8}.   
    \label{Invariants}
\end{align}
}
Here, $\ell=0,1/2,1,3/2,\ldots$ denotes the highest weight of $R_{2}$, and $(p,q)$ (with $p,q=0,1,2\ldots$) 
the weights of $R_{3}$. 

In the {\tt 210} approximation scheme, the $\beta$-functions are given by \cite{MachacekI,MachacekII,MachacekIII,Luo}
{\small
\bea
    \beta_{1}^{\mbox{\tiny NLO}} &=& \left(B_{1}+M_{1}\alpha_{1}+H_{1}\alpha_{2}+G_{1}\alpha_{3}-D_{1}\alpha_{y}-\frac{17}{3}\alpha_{t}\right)\alpha_{1}^{2},
   \nn  \label{Betaone} \\
    \beta_{2}^{\mbox{\tiny NLO}} &=& \Big(-B_{2}+M_{2}\alpha_{2}+H_{2}\alpha_{1}+G_{2}\alpha_{3}-D_{2}\alpha_{y}-3\, \alpha_{t}\Big)\alpha_{2}^{2},
 \nn    \label{Betatwo} \\
    \beta_{3}^{\mbox{\tiny NLO}} &=& \Big(-B_{3}+M_{3}\alpha_{3}+H_{3}\alpha_{1}+G_{3}\alpha_{2}-D_{3}\alpha_{y}-4\, \alpha_{t}\Big)\alpha_{3}^{2},
\nn     \label{Betathree} \\
    \beta_{t}^{\mbox{\tiny LO}} &=& \left(9\alpha_{t}-\frac{17}{6}\alpha_{1}-\frac{9}{2}\alpha_{2}-16\, \alpha_{3}\right)\alpha_{t},
 \nn    \label{BetatI} \\
    \beta_{y}^{\mbox{\tiny LO}} &=&\Big(T\alpha_{y}-F_{1}\alpha_{1}-F_{2}\alpha_{2}-F_{3}\alpha_{3}\Big)\alpha_{y},
    \label{BetayI}
\eea
}
where we have included the gauge and matter contributions in the coefficients $B_{i}$, $M_{i}$, $H_{i}$, $G_{i}$ and $D_{i}$, for $i=1,2,3$. These coefficient are expressed in terms of $d_{R_{2}}$, $d_{R_{3}}$, $C_{R_{2}}$, $C_{R_{3}}$, $S_{R_{2}}$, $S_{R_{3}}$, $Y$ and $N_{f}$ as follows. For the diagonal and mixing gauge contributions to the gauge $\beta$-functions we have
{\small
\begin{align} 
    B_{1}&=\frac{41}{3}+\frac{8}{3}N_{f}Y^{2}d_{R_{2}}d_{R_{3}}, & M_{1}&=\frac{199}{9}+8Y^{4}N_{f}d_{R_{2}}d_{R_{3}}, \nn  \\
    H_{1}&=9+8Y^{2}N_{f}C_{R_{2}}d_{R_{2}}d_{R_{3}}, & G_{1}&=\frac{88}{3}+8N_{f}Y^{2}C_{R_{3}}d_{R_{2}}d_{R_{3}}, \nn \\
    B_{2}&=\frac{19}{3}-\frac{8}{3}N_{f}S_{R_{2}}d_{R_{3}}, & M_{2}&=\frac{35}{3}+4N_{f}S_{R_{2}}d_{R_{3}}\left(2\, C_{R_{2}}+\frac{20}{3}\right), \nn \\
     H_{2}&=3+8N_{f}Y^{2}S_{R_{2}}d_{R_{3}}, & G_{2}&=24+8N_{f}S_{R_{2}}C_{R_{3}}d_{R_{3}}, \nn \\
     B_{3}&=14-\frac{8}{3}N_{f}S_{R_{3}}d_{R_{2}}, & M_{3}&=-52+4N_{f}S_{R_{3}}d_{R_{2}}(2\,C_{R_{3}}+10), \nn \\
     G_{3}&=9+8N_{f}S_{R_{3}}C_{R_{2}}d_{R_{2}}, & H_{3}&=\frac{11}{3}+8N_{f}Y^{2}S_{R_{3}}d_{R_{2}}. 
\end{align}
}
For the Yukawa contribution to the gauge $\beta$-functions we have
\begin{equation}
   D_{1}=4N_{f}^{2}Y^{2}d_{R_{2}}d_{R_{3}} , \  D_{2}=\frac{1}{3}4N_{f}^{2}C_{R_{2}}d_{R_{2}}d_{R_{3}}, \ D_{3}=\frac{1}{8}4N_{f}^{2}C_{R_{3}}d_{R_{2}}d_{R_{3}},
\end{equation}
whereas the running of the new coupling $\alpha_{y}$ is characterized by the coefficients
\begin{equation}
    T=2(N_{f}+d_{R_{2}}C_{R_{3}}), \ \ F_{1}=12\, Y^{2}, \ F_{2}=12\, C_{R_{2}}, \ F_{3}=12\, C_{R_{3}}.
    \label{YukawaI}
\end{equation}
All the new contributions to the gauge couplings running are multiplied by $N_{f}$,
meaning that we can go back to the SM by taking the $N_{f}\rightarrow0$ limit. 

Due to the simplicity of the $\beta$-functions to this order in perturbation theory, we can find analytic solutions of the equations 
$\beta_{i}^{\mbox{\tiny NLO}}=\beta_{t}^{\mbox{\tiny LO}}=\beta_{y}^{\mbox{\tiny LO}}=0$
as functions of $Y,\ell,p,q$ and $N_{f}$. 
All these solutions are listed in Table \ref{tab:grand}
and can be split in two categories according to whether they depend on the hypercharge $Y$
or not. All the latter  have $\alpha^{*}_{1}=0$.

For the gauge couplings, the $\beta$-functions in the {\tt 321} approximation scheme, are given, 
using the variables in \eq{couplings}, as follows
%
%
\bea
\beta_{1}^{\mbox{\tiny NNLO}}&=&\beta_{1}^{\mbox{\tiny NLO}}+
\left[
-M_{11}\alpha_{1}^{2}+M_{12}\alpha_{1}\alpha_{2}-M_{13}\alpha_{1}\alpha_{3} 
-G_{23}\alpha_{2}\alpha_{3} \vphantom{\frac{1}{2}} 
+H_{11}\alpha_{2}^{2}+G_{11}\alpha_{3}^{2}
\right.\nn \\
   & & \left.
+\frac{315}{8}\alpha_{t}^{2}+K_{y1}\alpha_{y}^{2}-\frac{2827}{144}\alpha_{1}\alpha_{t}-\frac{785}{16}\alpha_{2}\alpha_{t}-\frac{58}{3}\alpha_{3}\alpha_{t} \right.\nonumber \notag \\
&& \left.\vphantom{\frac{1}{2}} 
-\left(K_{11}\alpha_{1}+K_{12}\alpha_{2}+K_{13}\alpha_{3}\right)\alpha_{y}+\frac{3}{2}\left(\alpha_{1}+\alpha_{2}-\alpha_{\lambda}\right)\alpha_{\lambda}\right]\alpha_{1}^{2},
\nn    \label{BetaoneII} \\
\beta_{2}^{\mbox{\tiny NNLO}} &=&\beta_{2}^{\mbox{\tiny NLO}}+
   \left[ \vphantom{\frac{1}{2}}
-M_{22}\alpha_{2}^{2}+M_{21}\alpha_{2}\alpha_{1}-M_{23}\alpha_{2}\alpha_{3} 
-G_{13}\alpha_{1}\alpha_{3} \vphantom{\frac{1}{2}} 
-H_{22}\alpha_{1}^{2}+G_{22}\alpha_{3}^{2}
\right.\nn \\
& & \left.
+\frac{147}{8}\alpha_{t}^{2}+K_{y2}\alpha_{y}^{2}-\frac{729}{16}\alpha_{2}\alpha_{t}-\frac{593}{48}\alpha_{1}\alpha_{t}-14\, \alpha_{3}\alpha_{t} \right.\nonumber \notag \\
& & \left.\vphantom{\frac{1}{2}} 
-\left(K_{22}\alpha_{2}+K_{21}\alpha_{1}+K_{23}\alpha_{3}\right)\alpha_{y}+\frac{1}{2}\left(\alpha_{1}+3\, \alpha_{2}-3\, \alpha_{\lambda}\right)\alpha_{\lambda}\right]\alpha_{2}^{2},
\label{BetatwoII} \\
\beta_{3}^{\mbox{\tiny NNLO}} &=&\beta_{3}^{\mbox{\tiny NNLO}}+
\left[
-M_{33}\alpha_{3}^{2}+M_{31}\alpha_{3}\alpha_{1}-M_{32}\alpha_{3}\alpha_{2} 
-G_{12}\alpha_{1}\alpha_{2} \vphantom{\frac{1}{2}} 
-H_{33}\alpha_{2}^{2}+G_{33}\alpha_{2}^{2}
\right.\nn \\
& & \left.
+30\, \alpha_{t}^{2}
+K_{3y}\alpha_{y}^{2}-80\, \alpha_{3}\alpha_{t}-\frac{101}{12}\alpha_{1}\alpha_{t}-\frac{93}{4}\alpha_{2}\alpha_{t} 
\right.\nn \\
& & \left.\vphantom{\frac{1}{2}} 
-\left(K_{33}\alpha_{3}+K_{31}\alpha_{1}+K_{32}\alpha_{2}\right)\alpha_{y}
\right]\alpha_{3}^{2} \, .\label{BetathreeII} \nn
\eea
For the Yukawa and quartic Higgs couplings, the $\beta$-functions are given by
\bea
\beta_{t}^{\mbox{\tiny NLO}} &=&\beta_{t}^{\mbox{\tiny LO}}+
\left[
-24\, \alpha_{t}^{2}+3\, \alpha_{\lambda}^{2}-12\, \alpha_{t}\alpha_{\lambda}+\left(\frac{131}{8}\alpha_{1}+\frac{225}{8}\alpha_{2}+72\, \alpha_{3}\right)\alpha_{t} \right.\nonumber \notag \\
& &\left.\vphantom{\frac{1}{2}} +\frac{1187}{108}\alpha_{1}^{2}+\frac{3}{2}\alpha_{1}\alpha_{2}-\frac{23}{2}\alpha_{2}^{2}+\frac{38}{9}\alpha_{1}\alpha_{3}+18\, \alpha_{2}\alpha_{3}-216\, \alpha_{3}^{2} \right.\nonumber \notag \\
& & \left.\vphantom{\frac{1}{2}} +\frac{58}{27}B_{t1}\alpha_{1}^{2}+2B_{t2}\alpha_{2}^{2}+\frac{160}{9}B_{t3}\alpha_{3}^{2}\right]\alpha_{t}
 \label{BetatII} \\
\beta_{y}^{\mbox{\tiny NLO}} &=&\beta_{y}^{\mbox{\tiny LO}}+
\left[
(4-V)\alpha_{y}^{2}
+\left(V_1\alpha_{1}+V_2\alpha_{2}+V_3\alpha_{3}\right)\alpha_{y} 
\right.\nn \\
& & \Big.\vphantom{1} 
+W_{1}\alpha_{1}^{2}+W_{2}\alpha_{2}^{2}+W_{3}\alpha_{3}^{2}-W_{12}\alpha_{1}\alpha_{2}-W_{13}\alpha_{1}\alpha_{2}-W_{23}\alpha_{2}\alpha_{3}\Big]\alpha_{y},
\nn \label{BetayII} \\
\beta_{\lambda}^{\mbox{\tiny LO}} &=& 
12\, \alpha_{\lambda}^{2}-\left(3\, \alpha_{1}+9\, \alpha_{2}\right)\alpha_{\lambda}
+\frac{9}{4}\left(\frac{1}{3}\alpha_{1}^{2}
+\frac{2}{3}\alpha_{1}\alpha_{2}+\alpha_{2}^{2}\right)
+12\, \alpha_{t}\alpha_{\lambda}-12\, \alpha_{t}^{2}\, .
\nn \label{Betalambda}
\eea
In \eqs{BetatwoII}{BetatII}, we have introduced several coefficients containing the gauge and Yukawa contributions which depend on $N_f$ and the group representations of the SM and new vector-like fermions.
These coefficients are given in appendix \ref{sec:appcoff}.

It is not possible to find analytic solutions for the \FPs in the {\tt 321} approximation scheme. The system 
$\beta_{i} ^{\mbox{\tiny NNLO}}=\beta_{t}^{\mbox{\tiny NLO}}
=\beta_{y}^{\mbox{\tiny NLO}}= \beta_{\lambda}^{\mbox{\tiny LO}}=0$
must be solved numerically,
separately for each given choice of $(N_{f},Y,p,q,\ell)$. 
No separation between $Y$-independent and dependent solutions can be established before solving the equations.

\subsection{Results}
\label{sec:res}

In order to find \FPs satisfying the conditions (\ref{def}) and (\ref{thetabound}), we generate a grid in the space spanned by the quantum numbers 
$(N_{f},\ell,Y)$ for three specific $SU_c(3)$ representations: colorless ($p=q=0$), fundamental ($p=1, q=0$) and adjoint ($p=q=1$).
For each of these representations, we consider the following values
for the number of vector-like fermions, their isospin and hypercharge: 
$N_{f}\in[1,300]$ in steps of size 1, $\ell\in[1/2,10]$ and $Y\in[0,10]$ both in steps of size $1/2$. 
This amounts to 126,000 points for each representation of $SU_c(3)$. 

\subsubsection{Colorless vector-like fermions}

Colorless vector-like fermions are the least phenomenologically restricted and therefore the most attractive candidates for a successful 
extension of the SM. 
In the {\tt 210} approximation scheme we find that only the $Y$-independent set of solutions contains \FPs fulfilling 
the required conditions ($\alpha<1$, $|\vartheta|<O(1)$). 

\begin{figure}[h]
\includegraphics[width=\textwidth]{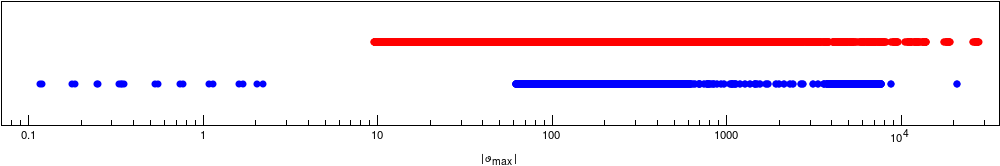}
\caption{\small \setstretch{1.0} Distribution of the largest eigenvalues $\vartheta_{\rm max}$ of the stability matrix of the \FPs of the colorless models. 
Blue dots: eigenvalues for the $Y$-independent solutions: there is a gap between 2.21 and 62.6.
Red dots: eigenvalues for the $Y$-dependent solutions: there is no gap, the eigenvalues start around 10.}
\label{Gapsinglet}
\end{figure}

To set the precise bound on $|\vartheta|$, we plot in Figure~\ref{Gapsinglet}
the largest eigenvalues of the stability matrix.
For the $Y$-independent solutions there is a gap between
$2.21$ and $62.6$; 
for the $Y$-dependent solutions there are no eigenvalues less than $9.63$.
Accordingly, we decide to consider \FPs with $|\vartheta|<3$.
In this way we probably include some \FPs that are not
within perturbation theory,but we prefer to err on this side
than to miss potentially interesting fixed points. 
In this way we discard all the $Y$-dependent \FPs since there is always an eigenvalue which is at least of order 10. 

After having applied all the criteria discussed in section \ref{sec:1} we find that,
for any value of the hypercharge $Y$, the only representations   producing satisfactory candidate \FPs are those collected, together with the corresponding eigenvalues,  in Table~\ref{tab:colorlessII}. 
The eigenvalues of the stability matrix turn out to be $Y$-independent as well. 
We also show in  Table~\ref{tab:colorlessII} the ratio $\rho_2$.
As discussed in section \ref{sec:tests},
this shows how large the three-loop contribution is with 
respect to the two-loop contribution. 

\begin{table}[h]
    \centering
{\scriptsize  \setstretch{1.5}
\begin{tabular}{|C|C|C|C|C|C|C|C|C|C|C|C|C|}\hline
    \rowcolor{Yellow}
    (N_{f},\ell) & \alpha^{*}_{1} & \alpha^{*}_{2} & \alpha^{*}_{3} & \alpha^{*}_{t} & \alpha^{*}_{y} & \vartheta_{1} & \vartheta_{2} & \vartheta_{3} & \vartheta_{4} & \vartheta_{5} &  & \rho_2\\ \hline
     \cellcolor{Yellow} (1,\frac{1}{2}) & 0 & 0.200 & 0 & 0 & 0.300 & 2.04 & -0.900 & 0.884 & 0 & 0 & P_{16} & 3.97 \\  
       \cellcolor{Yellow} & 0 & 0.213 & 0 & 0.106 & 0.319 & 2.21 & 1.19 & 0.743 & 0 & 0 & P_{17} & 4.33\\  
         \cellcolor{Yellow}& 0 & 0.179 & 0 & 0 & 0 & -1.61 & 0.893 & -0.804 & 0 & 0 & P_{18} & 3.28\\  
           \cellcolor{Yellow} & 0 & 0.189 & 0 & 0.0943 & 0 & -1.70 & 1.15 & 0.697 & 0 & 0 & P_{19} & 3.53 \\  \hline
   \cellcolor{Yellow} (1,1) & 0 & 0.0137 & 0 & 0 & 0.0411 & 0.333 & -0.0616 & 0.0135 & 0 & 0 & P_{16} & 0.194 \\  
   \rowcolor{Green!20}
\cellcolor{Yellow}& 0 & 0.0140 & 0 & 0.0070 & 0.0420 & 0.341 & 0.0633 & 0.0137 & 0 & 0 & P_{17} & 0.198\\ 
 \cellcolor{Yellow}& 0 & 0.0103 & 0 & 0 & 0 & -0.247 & -0.0464 & 0.0103 & 0 & 0 & P_{18} & 0.0963 \\ 
 \rowcolor{Green!20}
 \cellcolor{Yellow}& 0 & 0.0105 & 0 & 0.0052 & 0 & -0.251 & 0.0473 & 0.0104 & 0 & 0 & P_{19} & 0.0973 \\
 \hline
     \cellcolor{Yellow} (2,\frac{1}{2}) & 0 & 0.104 & 0 & 0 & 0.117 & 1.0833 & -0.467 & 0.328 & 0 & 0 & P_{16} & 1.71 \\  
     \rowcolor{Green!20}
       \cellcolor{Yellow} & 0 & 0.108 & 0 & 0.0542 & 0.122 & 1.14 & 0.525 & 0.315 & 0 & 0 & P_{17} & 1.81 \\
\cellcolor{Yellow} & 0 & 0.0827 & 0 & 0 & 0 & -0.744 & -0.372 & 0.303 & 0 & 0 & P_{18} & 1.19 \\ 
\rowcolor{Green!20}
 \cellcolor{Yellow}& 0 & 0.0856 & 0 & 0.0428 & 0 & -0.770 & 0.427 & 0.283 & 0 & 0 & P_{19} & 1.23 \\ 
 \hline
 \cellcolor{Yellow}(3,\frac{1}{2}) & 0 & 0.0525 & 0 & 0 & 0.0472 & 0.530 & -0.236 & 0.109 & 0 & 0 & P_{16} & 0.763 \\ 
  \rowcolor{Green!20}
\cellcolor{Yellow}& 0 & 0.0543 & 0 & 0.0272 & 0.0489 & 0.552 & 0.251 & 0.109 & 0 & 0 & P_{17} & 0.794 \\
 \cellcolor{Yellow}& 0 & 0.0385 & 0 & 0 & 0 & -0.346 & -0.173 & 0.0897 & 0 & 0 & P_{18} & 0.471 \\ 
\rowcolor{Green!20}
 \cellcolor{Yellow}& 0 & 0.0394 & 0 & 0.0197 & 0 & -0.355 & 0.182 & 0.0896 & 0 & 0 & P_{19} & 0.483 \\  
 \hline
 \cellcolor{Yellow}(4,\frac{1}{2}) & 0 & 0.0189 & 0 & 0 & 0.0141 & 0.179 & -0.0849 & 0.0179 & 0 & 0 & P_{16} & 0.246 \\ 
 \rowcolor{Green!20}
 \cellcolor{Yellow}& 0 & 0.0194 & 0 & 0.0097 & 0.0146 & 0.185 & 0.0880 & 0.0182 & 0 & 0 & P_{17} & 0.253 \\ 
 \cellcolor{Yellow}& 0 & 0.0130 & 0 & 0 & 0 & -0.117 & -0.0584 & 0.0130 & 0 & 0 & P_{18} & 0.141 \\ 
\rowcolor{Green!20}
 \cellcolor{Yellow}& 0 & 0.0132 & 0 & 0.0066 & 0 & -0.119 & 0.0599 & 0.0132 & 0 & 0 & P_{19} & 0.143 \\ 
    \hline
   \end{tabular}
   }
   \caption{\small  \setstretch{1.0} Set of  \FPs and eigenvalues for colorless vector-like fermions in the {\tt 210} approximation scheme. 
   We highlight in green the \FPs that appear also in the {\tt 321} approximation. 
   The labels in the second to the last last column refer to the list in Table \ref{tab:grand}.
   We show in the last column the ratio $\rho_2$  defined in Eq. (\ref{LoopsFP}) knowing that in {\tt 210} $A_{*}^{(2)}=B_{*}^{(2)}$.}
   \label{tab:colorlessII}
   \end{table}

\begin{figure}[h]
\centering
\includegraphics[scale=0.56]{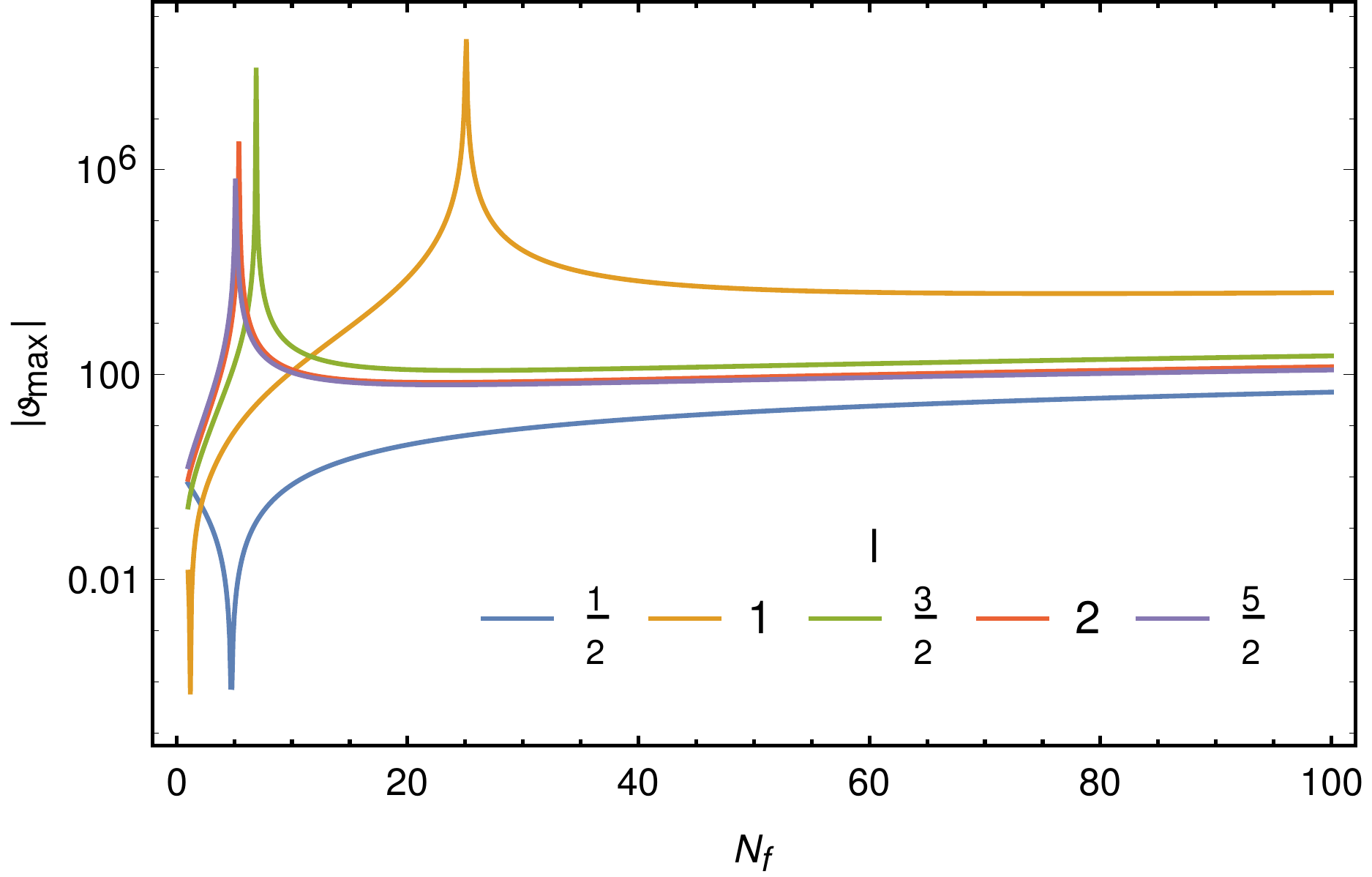}
\caption{\small \setstretch{1.0} Behaviour of a given eigenvalue $|\vartheta|$ as a function of $N_{f}$ for several values of $\ell$ in the colorless case. 
The  scaling dimension increases very fast with $N_{f}$, and only small values of $N_{f},\ell$ 
produce $|\vartheta|<O(1)$.}
\label{fig:EigvalsNTL}
\end{figure}

The bounds on $N_{f}$ and $\ell$ come from the behavior of the eigenvalues as functions of these parameters. If we plot one of the eigenvalues as a 
function of $N_{f}$ for several values of $l$, we observe that it increases very fast. From Figure~\ref{fig:EigvalsNTL}, we see that only models with 
small $N_{f}$ produce sufficiently small eigenvalues. 

It is important to note that the large scaling dimensions of  models with large $N_f$ frustrate  the apparently promising strategy of increasing $N_f$ in order to increase the NLO term in the gauge $\beta$-functions to cancel the ($N_f$-independent) 
LO term with smaller (and therefore more perturbative) values of the couplings $\alpha_i$.

The above selection of the viable \FPs  is  confirmed by the study of their CFT central charges. 
There are 20 $Y$-independent \FPs with eigenvalues up to about $\pm 2$.
The \FP with least variation in the central charges is that with $(N_f, \ell)=(1,1)$, having $\delta a \simeq -0.0007$ and $\delta c \simeq 0.08$. The one with the largest change  is that with $(N_f, \ell)=(1, 1/2)$, having $\delta a \simeq -0.2$ and $\delta c \simeq 0.8$. All these \FPs (except for the one corresponding to $(N_f, \ell, Y) = (1,1/2,0)$) pass the collider bounds test (see appendix \ref{sec:CFT}). 
There are 69 $Y$-dependent \FPs with eigenvalues up to $\pm10$. None of them have positive $a$ or $c$ with $\delta a$ and $\delta c$ being of  $O(1)$. They should all be discarded. These results confirm  our  classification of the \FPs   in Table \ref{tab:colorlessII} according to the size of their eigenvalues and the ratio $\rho$.

Now that we have isolated the candidates to study, we check whether these \FPs can be connected to the SM via the RG flow.  
We find that $\beta_{1}$ is proportional to $\alpha_{1}^{2}$
and so, in order to avoid Landau poles, 
$\alpha_{1}$ has to vanishes at all energy scales. 
In conclusion, although we have perturbative fixed points, 
these cannot be matched to the SM because we know that $g_{1}$ 
is different from zero at the TeV scale.

We then perform a similar search in the {\tt 321} approximation scheme. Since we see in Table \ref{tab:colorlessII} that the \FP with $|\vartheta|>1$
produce a rather large $\rho_2$ ratio, we stick to solutions having $|\vartheta|<1$.
We find that the same combinations of $N_{f}$ and $\ell$ that provide perturbative \FPs in the {\tt 210} case 
also give viable solutions here. 
Moreover, the solutions turn out to be $Y$-independent as well. 

In Table \ref{ColorlessNNLO} we show the \FP solutions satisfying  the criteria in \eq{def} and \eq{thetabound}.
All the \FPs in Table \ref{ColorlessNNLO} can be traced back to \FPs that were already
present in the {\tt 210} approximation scheme and listed in Table ~\ref{tab:colorlessII}. 
Notice that  for a given pair $(N_{f},\ell)$, not all the \FPs in {\tt 210} persist.
For those that do, the values of $\alpha^{*}$ and $\vartheta$
change by relatively small amount. 
We can then claim that the solutions given in Table \ref{ColorlessNNLO} are radiatively stable fixed points.

\begin{table}[h]
    \centering
   { \scriptsize  \setstretch{1.5} 
    \begin{tabular}{|C|C|C|C|C|C|C|C|C|C|C|C|C|C|C|}\hline
    \rowcolor{Yellow}
     (N_{f},l) & \alpha^{*}_{1} & \alpha^{*}_{2} & \alpha^{*}_{3} & \alpha^{*}_{t} & \alpha^{*}_{y} & \alpha^{*}_{\lambda} & \vartheta_{1} & \vartheta_{2} & \vartheta_{3} & \vartheta_{4} & \vartheta_{5} & \vartheta_{6} & \sigma_2 & \rho_2\\ 
\hline
\rowcolor{Green!20}
\cellcolor{Yellow} (1,1) & 0 & 0.0096 & 0 & 0.0048 & 0 & 0.0039 & -0.244 & 0.0655 & 0.0430 & 0.0103 & 0 & 0 & 0.918 & 0.0821 \\ \rowcolor{Green!20}
 \cellcolor{Yellow} &  0 & 0.0119 & 0 & 0.0060 & 0.0343 & 0.0048 & 0.301 & 0.0813 & 0.0531 & 0.0134 & 0 & 0 & 0.8601 & 0.140 \\  
 \hline\rowcolor{Green!20}
  \cellcolor{Yellow}(2,\frac{1}{2}) & 0 & 0.0498 & 0 & 0.0259 & 0 & 0.0211 & -0.592 & 0.382 & 0.282 & 0.200 & 0 & 0 & 0.581 & 0.418 \\ \rowcolor{Green!20}
 \cellcolor{Yellow}& 0 & 0.0567& 0 & 0.0296 & 0.0734 & 0.0242 & 0.696 & 0.442 & 0.314 & 0.224 & 0 & 0 & 0.5012 & 0.499 \\ 
 \hline\rowcolor{Green!20}
 \cellcolor{Yellow}(3,\frac{1}{2}) & 0 & 0.0291 & 0 & 0.0148 & 0 & 0.0120 & -0.306 & 0.2080 & 0.132 & 0.0827 & 0 & 0 & 0.737 & 0.263 \\ \rowcolor{Green!20}
 \cellcolor{Yellow}& 0 & 0.0362 & 0 & 0.0184 & 0.0353 & 0.0150 & 0.403 & 0.262 & 0.165 & 0.100 & 0 & 0 & 0.645 & 0.354 \\ 
 \hline\rowcolor{Green!20}
 \cellcolor{Yellow}(4,\frac{1}{2}) & 0 & 0.0117 & 0 & 0.0059 & 0 & 0.0048 & -0.112 & 0.0804 & 0.052 & 0.0130 & 0 & 0 & 0.887 & 0.113 \\ \rowcolor{Green!20}
 \cellcolor{Yellow}& 0 & 0.0162 & 0 & 0.0081 & 0.0125 & 0.0066 & 0.161 & 0.112 & 0.0723 & 0.0179 & 0 & 0 & 0.823 & 0.177 \\ 
    \hline
   \end{tabular}
   }
\caption{\small \setstretch{1.0} Fixed points and eigenvalues for colorless vector-like fermions,
in the {\tt 321} approximation scheme.  
The last two columns give the values of the ratios $\sigma_2$ and $\rho_2$ (see \ref{LoopsFPbound}).}
\label{ColorlessNNLO}
\end{table}

Unfortunately, when we look at trajectories lying on the UV critical surface, 
we find  again that the coupling $\alpha_{1}$ must be zero at all scales in all the models. 
The abelian interactions suffer from the triviality problem and no matching to the SM is  possible if asymptotic safety is assumed. 

All these colorless models are therefore ruled out.

\subsubsection{Vector-like fermions in the fundamental of $SU_c(3)$} 

\begin{figure}[H]
\includegraphics[width=\textwidth]{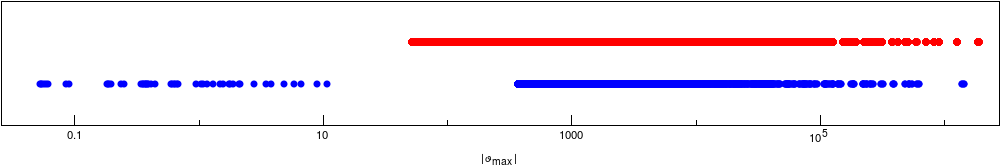}
\caption{\small \setstretch{1.0} Distribution of the largest eigenvalues $\vartheta_{\rm max}$ of the stability matrix of the \FPs of the $SU(3)$ fundamental representation. Blue dots: eigenvalues for the $Y$-independent solutions: there is a gap between 10.8 and 372.
Red dots: eigenvalues for the $Y$-dependent solutions: there is no gap, the eigenvalues start at 52.1.}
\label{Gapfund}
\end{figure}

For the fundamental representation ($p=1$ and $q=0$ or vice-versa) we follow the same procedure as before and generate 126,000 models by scanning the same grid in 
the $(N_f,\ell,Y)$ space. 
We split the solutions in two families depending on whether they  depend on the value of their hypercharge $Y$  or not. 
The distribution of the largest eigenvalues given
in Figure~\ref{Gapfund} shows that there are no \FPs with 
$|\vartheta|<52.1$ for the $Y$-dependent solutions, whereas for the 
$Y$-independent solutions there is a gap between $10.8$ and $372$. 
Accordingly, we eliminate all $Y$-dependent solutions and impose the bound 
$|\vartheta|<11$ for those that are  $Y$-independent.
In this way, even more than in the preceding section, we include
models that are probably 
\noindent
unreliable, but these can be eliminated at a later stage.
For the $Y$-independent solutions, we find the  combinations of 
$N_{f}$ and $\ell$ in Tables \ref{FundtableII} and \ref{FundtableIII} that generate satisfactory candidate fixed points.

This  selection is confirmed by the study of the central charges for these models.
Among the 49 distinct $Y$-independent \FPs with eigenvalues up to $\pm10$, all have positive $c$-function, but 6 of them have a negative $a$-function (with one more being borderline acceptable). The CFT test seems to work well here: all \FPs with reasonable critical exponents pass it, whereas the ones with relatively large exponents do not. 
An unexpected fact is that the separation between large and small exponents seems to be around a maximum value of $|\vartheta|$ around 3. For these perturbative and ``semi-perturbative''  fixed points, we also notice that the $a$-function is generically pushed toward 0 ($a_q<0$) whereas the $c$ function is generically shifted to larger values ($c_q>0$). This is why the \FPs with negative $a$-function still seem to pass the $c$-function test. If one considers $\delta c$ instead, then for most of these \FPs $\delta c>1$, but apparently not for all. Finally, if one also studies the collider bounds one finds that ten more \FPs are excluded, usually those which just barely satisfied one or both of the $a$ and $c$ tests. The collider bounds tests seem to be the most stringent.
    
When one tries to match these \FPs to the SM at low energies,
it turns out that the abelian gauge coupling $\alpha_1$ 
must again be zero at all scales.
None of these \FPs is physically viable.

In the {\tt 321} approximation scheme, there exist \FPs 
that can be reasonably traced back to those in the 
{\tt 210} approximation scheme. 
These solutions are shown in Table \ref{FundNNLO}, where we have included only \FPs with $|\vartheta|<1$ in order to get small ratios $\rho_i$ and $\sigma_i$.
However, they all have at least one coupling that has to be zero at all scales,
thus preventing a proper matching to the SM. 

We conclude that also all the models with the vector-like fermions in the fundamental representation of $SU_c(3)$ cannot provide an AS extension to the SM. 

\begin{table}[h]
\centering
\scriptsize
\begin{tabular}{|C|C|C|C|C|C|C|C|C|C|C|C|C|}\hline
\rowcolor{Yellow}
   \cellcolor{Yellow} (N_{f},l) & \alpha^{*}_{1} & \alpha^{*}_{2} & \alpha^{*}_{3} & \alpha^{*}_{t} & \alpha^{*}_{y} & \vartheta_{1} & \vartheta_{2} & \vartheta_{3} & \vartheta_{4} & \vartheta_{5} &  & \rho\\ \hline
 \cellcolor{Yellow}   (1,\frac{1}{2}) & 0 & 0.0411 & 0 & 0 & 0.0264 & 0.378 & -0.185 & 0.0936 & 0 & 0 & P_{16} & 0.522 \\ 
  \rowcolor{Green!20}
 \cellcolor{Yellow}& 0 & 0.0422 & 0 & 0.0211 & 0.0271 & 0.389 & 0.195 & 0.0936 & 0 & 0 & P_{17} & 0.537 \\ 
 \cellcolor{Yellow}& 0 & 0.0385 & 0 & 0 & 0 & -0.346 & -0.173 & 0.0897 & 0 & 0 & P_{18} & 0.471 \\ \rowcolor{Green!20}
 \cellcolor{Yellow}& 0 & 0.0394 & 0 &
  0.0197 & 0 & -0.355 & 0.182 & 0.0896 & 0 & 0 & P_{19} & 0.483 \\
 \hline
  \cellcolor{Yellow}   (1,1) & 0 & 0 & 0.417 & 0 & 0 & -6.67 & -6.67 & 4.17 & 0 & 0 & P_{11} & 20.9 \\ 
   \cellcolor{Yellow}  & 0 & 0 & 0.521 & 0 & 0.417 & 10.8 & -8.33 & 4.00 & 0 & 0 & P_{9} & 31.8 \\  \hline
 \cellcolor{Yellow} (1,\frac{3}{2}) & 0 & 0 & 0.176 & 0 & 0 & -2.81 & -2.81 & 1.52 & 0 & 0 & P_{11} & 5.45 \\ 
 \cellcolor{Yellow}& 0 & 0 & 0.205 & 0.365 & 0 & 3.84 & -3.28 & 1.52 & 0 & 0 & P_{10} & 7.21 \\ 
 \cellcolor{Yellow}& 0 & 0 & 0.195 & 0 & 0.120 & 3.49 & -3.12 & 1.51 & 0 & 0 & P_{9} & 6.60 \\ 
 \cellcolor{Yellow}& 0 & 0 & 0.232 & 0.413 & 0.143 & 4.83 & 3.72 & 1.55 & 0 & 0 & P_{8} & 9.06 \\ \hline
 \cellcolor{Yellow} (1,2) & 0 & 0 & 0.0982 & 0 & 0 & -1.57 & -1.57 & 0.720 & 0 & 0 & P_{11} & 2.42 \\ 
 \cellcolor{Yellow}& 0 & 0 & 0.108 & 0.193 & 0 & 1.88 & -1.74 & 0.735 & 0 & 0 & P_{10} & 2.88 \\ 
 \cellcolor{Yellow}& 0 & 0 & 0.105 & 0 & 0.0526 & 1.78 & -1.68 & 0.730 & 0 & 0 & P_{9} & 2.73 \\ 
 \cellcolor{Yellow}& 0 & 0 & 0.117 & 0.208 & 0.0586 & 2.15 & 1.88 & 0.749 & 0 & 0 & P_{8} & 3.30 \\ \hline
 \rowcolor{Green!20}
 \cellcolor{Yellow}(1,\frac{5}{2}) & 0 & 0 & 0.0600 & 0 & 0 & -0.960 & -0.960 & 0.360 & 0 & 0 & P_{11} & 1.27 \\ 
 \cellcolor{Yellow}& 0 & 0 & 0.0646 & 0.115 & 0 & 1.08 & -1.03 & 0.371 & 0 & 0 & P_{10} & 1.44 \\ \rowcolor{Green!20}
 \cellcolor{Yellow}& 0 & 0 & 0.0632 & 0 & 0.0266 & 1.04 & -1.01 & 0.368 & 0 & 0 & P_{9} & 1.39 \\ 
 \cellcolor{Yellow}& 0 & 0 & 0.0683 & 0.121 & 0.0288 & 1.18 & 1.09 & 0.380 & 0 & 0 & P_{8} & 1.59 \\
 \hline
 \cellcolor{Yellow}   (1,3) & 0 & 0 & 0.0412 & 0.0733 & 0.0150 & 0.689 & 0.660 & 0.184 & 0 & 0 & P_{8} & 0.839 \\ 
  \rowcolor{Green!20}
 \cellcolor{Yellow}& 0 & 0 & 0.0388 & 0 & 0.0141 & 0.632 & -0.621 & 0.178 & 0 & 0 & P_{9} & 0.758 \\
 \cellcolor{Yellow} & 0 & 0 & 0.0395 & 0.0702 & 0 & 0.647 & -0.632 & 0.180 & 0 & 0 & P_{10} & 0.778 \\ \rowcolor{Green!20}
 \cellcolor{Yellow}& 0 & 0 & 0.0372 & 0 & 0 & -0.596 & -0.596 & 0.174 & 0 & 0 & P_{11} & 0.707 \\ 
 \hline\rowcolor{Green!20}
 \cellcolor{Yellow} (1,\frac{7}{2}) & 0 & 0 & 0.0221 & 0 & 0 & -0.354 & -0.354 & 0.0737 & 0 & 0 & P_{11} & 0.384 \\ \rowcolor{Green!20}
 \cellcolor{Yellow}& 0 & 0 & 0.0232 & 0.0413 & 0 & 0.376 & -0.371 & 0.0764 & 0 & 0 & P_{10} & 0.415 \\ \rowcolor{Green!20}
 \cellcolor{Yellow}& 0 & 0 & 0.0229 & 0 & 0.0073 & 0.370 & -0.366 & 0.0756 & 0 & 0 & P_{9} & 0.406 \\ \rowcolor{Green!20}
 \cellcolor{Yellow}& 0 & 0 & 0.0241 & 0.0428 & 0.0077 & 0.394 & 0.385 & 0.0784 & 0 & 0 & P_{8} & 0.441 \\
 \hline\rowcolor{Green!20}
   \cellcolor{Yellow} (1,4) & 0 & 0 & 0.0114 & 0 & 0 & -0.182 & -0.182 & 0.0235 & 0 & 0 & P_{11} & 0.182 \\ \rowcolor{Green!20}
 \cellcolor{Yellow}& 0 & 0 & 0.0118 & 0.0210 & 0 & 0.191 & -0.189 & 0.0235 & 0 & 0 & P_{10} & 0.195 \\ \rowcolor{Green!20}
 \cellcolor{Yellow}& 0 & 0 & 0.0117 & 0 & 0.0033 & 0.188 & -0.187 & 0.0233 & 0 & 0 & P_{9} & 0.191 \\ \rowcolor{Green!20}
 \cellcolor{Yellow}& 0 & 0 & 0.0122 & 0.0217 & 0.0035 & 0.197 & 0.195 & 0.0242 & 0 & 0 & P_{8} & 0.205 \\ \rowcolor{Green!20}
 \hline
 \cellcolor{Yellow}(1,\frac{9}{2}) & 0 & 0 & 0.0033 & 0 & 0 & -0.0530 & -0.0530 & 0.0022 & 0 & 0 & P_{11} & 0.0495 \\ \rowcolor{Green!20}
 \cellcolor{Yellow}& 0 & 0 & 0.0034 & 0.0061 & 0 & 0.0550 & -0.0549 & 0.0023 & 0 & 0 & P_{10} & 0.0523 \\ \rowcolor{Green!20}
 \cellcolor{Yellow}& 0 & 0 & 0.0034 & 0 & 0.0009 & 0.0544 & -0.0544 & 0.0023 & 0 & 0 & P_{9} & 0.0516 \\ \rowcolor{Green!20}
 \cellcolor{Yellow}& 0 & 0 & 0.0035 & 0.0063 & 0.0009 & 0.0566 & 0.0564 & 0.0023 & 0 & 0 & P_{8} & 0.0547 \\ \rowcolor{Green!20}
 \hline
\end{tabular}
\caption{\small \setstretch{1.0} Fixed points and eigenvalues for vector-like fermions in the fundamental representation of $SU_c(3)$,
in the {\tt 210} approximation scheme, with $N_f=1$.  
We highlight in green the \FPs that appear also in the {\tt $321$} approximation scheme. 
The labels in the second to the last last column refer to the list in Table \ref{tab:grand}.
The last column gives the values of the ratio $\rho$ for $\alpha_2$ or $\alpha_3$ depending on the case (see \ref{LoopsFPbound}).
}
\label{FundtableII}
\end{table}

\begin{table}[h]
\centering
\scriptsize
\begin{tabular}{|C|C|C|C|C|C|C|C|C|C|C|C|C|}\hline
\rowcolor{Yellow}
   \cellcolor{Yellow} (N_{f},l) & \alpha^{*}_{1} & \alpha^{*}_{2} & \alpha^{*}_{3} & \alpha^{*}_{t} & \alpha^{*}_{y} & \vartheta_{1} & \vartheta_{2} & \vartheta_{3} & \vartheta_{4} & \vartheta_{5} &  & \rho\\ \hline
 \cellcolor{Yellow}(2,\frac{1}{2}) & 0 & 0 & 0.176 & 0 & 0 & -2.81 & -2.81 & 1.52 & 0 & 0 & P_{11} & 5.45 \\ 
 \cellcolor{Yellow}& 0 & 0 & 0.205 & 0.365 & 0 & 3.84 & -3.28 & 1.52 & 0 & 0 & P_{10} & 7.21 \\ 
 \cellcolor{Yellow}& 0 & 0 & 0.260 & 0 & 0.260 & 5.91 & -4.16 & 1.59 & 0 & 0 & P_{9} & 11.1 \\ 
 \cellcolor{Yellow}& 0 & 0 & 0.330 & 0.588 & 0.330 & 8.99 & 5.29 & 1.68 & 0 & 0 & P_{8} & 17.4 \\
 \hline
   \rowcolor{Green!20}
    \cellcolor{Yellow}(2,1) & 0 & 0 & 0.0600 & 0 & 0 & -0.960 & -0.960 & 0.360 & 0 & 0 & P_{11} & 1.27 \\  
 \cellcolor{Yellow}& 0 & 0 & 0.0646 & 0.115 & 0 & 1.08 & -1.03 & 0.371 & 0 & 0 & P_{10} & 1.44 \\ \rowcolor{Green!20}
 \cellcolor{Yellow}& 0 & 0 & 0.0727 & 0 & 0.0529 & 1.30 & -1.16 & 0.390 & 0 & 0 & P_{9} & 1.77 \\ 
 \cellcolor{Yellow}& 0 & 0 & 0.0795 & 0.141 & 0.0578 & 1.50 & 1.27 & 0.405 & 0 & 0 & P_{8} & 2.07 \\
 \hline\rowcolor{Green!20}
 \cellcolor{Yellow}(2,\frac{3}{2}) & 0 & 0 & 0.0221 & 0 & 0 & -0.354 & -0.354 & 0.0737 & 0 & 0 & P_{11} & 0.384 \\ \rowcolor{Green!20}
 \cellcolor{Yellow}& 0 & 0 & 0.0232 & 0.0413 & 0 & 0.376 & -0.371 & 0.0764 & 0 & 0 & P_{10} & 0.415 \\ \rowcolor{Green!20}
 \cellcolor{Yellow}& 0 & 0 & 0.0252 & 0 & 0.0144 & 0.417 & -0.403 & 0.0810 & 0 & 0 & P_{9} & 0.475 \\
 \cellcolor{Yellow}& 0 & 0 & 0.0266 & 0.0473 & 0.0152 & 0.448 & 0.426 & 0.0842 & 0 & 0 & P_{8} & 0.520 \\
 \hline\rowcolor{Green!20}
 \cellcolor{Yellow}   (2,2) & 0 & 0 & 0.0033 & 0 & 0 & -0.0530 & -0.0530 & 0.0022 & 0 & 0 & P_{11} & 0.0495 \\ \rowcolor{Green!20}
 \cellcolor{Yellow}& 0 & 0 & 0.0034 & 0.0061 & 0 & 0.0550 & -0.0549 & 0.0023 & 0 & 0 & P_{10} & 0.0523 \\ \rowcolor{Green!20}
 \cellcolor{Yellow}& 0 & 0 & 0.0036 & 0 & 0.0017 & 0.0587 & -0.0584 & 0.0024 & 0 & 0 & P_{9} & 0.0579 \\ \rowcolor{Green!20}
 \cellcolor{Yellow}& 0 & 0 & 0.0038 & 0.0068 & 0.0018 & 0.0612 & 0.0608 & 0.0025 & 0 & 0 & P_{8} & 0.0616 \\ 
 \hline\rowcolor{Green!20}
 \cellcolor{Yellow}(3,\frac{1}{2}) & 0 & 0 & 0.0600 & 0 & 0 & -0.960 & -0.960 & 0.360 & 0 & 0 & P_{11} & 1.27 \\ 
 \cellcolor{Yellow}& 0 & 0 & 0.0646 & 0.115 & 0 & 1.08 & -1.03 & 0.371 & 0 & 0 & P_{10} & 1.44 \\ \rowcolor{Green!20}
 \cellcolor{Yellow}& 0 & 0 & 0.0882 & 0 & 0.0784 & 1.77 & -1.41 & 0.423 & 0 & 0 & P_{9} & 2.47 \\ 
 \cellcolor{Yellow}& 0 & 0 & 0.0985 & 0.175 & 0.0876 & 2.10 & 1.58 & 0.443 & 0 & 0 & P_{8} & 3.01 \\
 \hline\rowcolor{Green!20}
 \cellcolor{Yellow}   (3,1) & 0 & 0 & 0.0114 & 0 & 0 & -0.182 & -0.182 & 0.0227 & 0 & 0 & P_{11} & 0.182 \\ \rowcolor{Green!20}
 \cellcolor{Yellow}& 0 & 0 & 0.0118 & 0.0210 & 0 & 0.191 & -0.189 & 0.0235 & 0 & 0 & P_{10} & 0.195 \\ \rowcolor{Green!20}
 \cellcolor{Yellow}& 0 & 0 & 0.0143 & 0 & 0.0095 & 0.237 & -0.229 & 0.0276 & 0 & 0 & P_{9} & 0.264 \\ \rowcolor{Green!20}
 \cellcolor{Yellow}& 0 & 0 & 0.0150 & 0.0267 & 0.0100 & 0.252 & 0.241 & 0.0288 & 0 & 0 & P_{8} & 0.288 \\ 
 \hline\rowcolor{Green!20}
 \cellcolor{Yellow}(4,\frac{1}{2})  & 0 & 0 & 0.0221 & 0 & 0 & -0.354 & -0.354 & 0.0737 & 0 & 0 & P_{11} & 0.384 \\ \rowcolor{Green!20}
 \cellcolor{Yellow}& 0 & 0 & 0.0232 & 0.0413 & 0 & 0.376 & -0.371 & 0.0764 & 0 & 0 & P_{10} & 0.415 \\ \rowcolor{Green!20}
 \cellcolor{Yellow} & 0 & 0 & 0.0335 & 0 & 0.0268 & 0.607 & -0536 & 0.0987 & 0 & 0 & P_{9} & 0.763 \\ 
 \cellcolor{Yellow}& 0 & 0 & 0.0361 & 0.0642 & 0.0289 & 0.670 & 0.577 & 0.104 & 0 & 0 & P_{8} & 0.866 \\ 
 \hline\rowcolor{Green!20}
 \cellcolor{Yellow}(5,\frac{1}{2}) & 0 & 0 & 0.0033 & 0 & 0 & -0.0530 & -0.530 & 0.0022 & 0 & 0 & P_{11} & 0.0495 \\ \rowcolor{Green!20}
 \cellcolor{Yellow}& 0 & 0 & 0.0343 & 0.0061 & 0 & 0.0550 & -0.0549 & 0.0023 & 0 & 0 & P_{10} & 0.0523 \\ \rowcolor{Green!20}
 \cellcolor{Yellow}& 0 & 0 & 0.0052 & 0 & 0.0038 & 0.0850 & -0.0829 & 0.0034 & 0 & 0 & P_{9} & 0.1010 \\ \rowcolor{Green!20}
 \cellcolor{Yellow}& 0 & 0 & 0.0055 & 0.0097 & 0.0040 & 0.0903 & 0.0878 & 0.035 & 0 & 0 & P_{8} & 0.111 \\
    \hline
\end{tabular}
\caption{\small \setstretch{1.0} Same as Table \ref{FundtableII} but with $N_f >1$.
}
\label{FundtableIII}
\end{table}

\begin{table}[h]
    \centering
\scriptsize 
\begin{tabular}{|C|C|C|C|C|C|C|C|C|C|C|C|C|C|C|}\hline\rowcolor{Yellow}
    (N_{f},l) & \alpha^{*}_{1} & \alpha^{*}_{2} & \alpha^{*}_{3} & \alpha^{*}_{t} & \alpha^{*}_{y} & \alpha^{*}_{\lambda} & \vartheta_{1} & \vartheta_{2} & \vartheta_{3} & \vartheta_{4} & \vartheta_{5} & \vartheta_{6} & \sigma & \rho \\ \hline\rowcolor{Green!20}
  \cellcolor{Yellow}  (1,\frac{1}{2}) & 0 & 0.0291 & 0 & 0.0148 & 0 & 0.0120& -0.306& 0.208 & 0.132 & 0.0827 & 0 & 0 & 0.737 & 0.263 \\ \rowcolor{Green!20}
 \cellcolor{Yellow}& 0 & 0.0305 & 0 & 0.0155 & 0.0209 & 0.0126 & 0.322 & 0.219 & 0.139 & 0.0863 & 0 & 0 & 0.719 & 0.281 \\ 
 \hline\rowcolor{Green!20}
 \cellcolor{Yellow}(1,\frac{5}{2}) & 0 & 0 & 0.0346 & 0 & 0 & 0 & -0.748 & -0.748 & 0.295 & 0 & 0 & 0 & 0.577 & 0.423 \\ \rowcolor{Green!20}
 \cellcolor{Yellow}& 0 & 0 & 0.0355 & 0 & 0.0167 & 0 & -0.774 & 0.768 & 0.304 & 0 & 0 & 0 & 0.559 & 0.441 \\ 
 \hline\rowcolor{Green!20}
 \cellcolor{Yellow}   (1,3) & 0 & 0 & 0.0252 & 0 & 0 & 0 & -0.501 & -0.501 & 0.156 & 0 & 0 & 0 & 0.676 & 0.323 \\ \rowcolor{Green!20}
 \cellcolor{Yellow}& 0 & 0 & 0.0258 & 0 & 0.0101 & 0 & -0.516 & 0.514 & 0.160 & 0 & 0 & 0 & 0.664 & 0.336 \\  
 \hline\rowcolor{Green!20}
 \cellcolor{Yellow}(1,\frac{7}{2}) & 0 & 0 & 0.0171 & 0 & 0 & 0 & -0.315 & -0.315 & 0.0670 & 0 & 0 & 0 & 0.771 & 0.228 \\ \rowcolor{Green!20}
 \cellcolor{Yellow}& 0 & 0 & 0.0177 & 0.0358 & 0 & 0.0221 & 0.969 & -0.329 & 0.290 & 0.0723 & 0 & 0 & 0.758 & 0.242 \\ \rowcolor{Green!20}
 \cellcolor{Yellow}& 0 & 0 & 0.0175 & 0 & 0.0058 & 0 & -0.324 & 0.324 & 0.0717 & 0 & 0 & 0 & 0.763 & 0.237 \\ \rowcolor{Green!20}
 \cellcolor{Yellow}& 0 & 0 & 0.0182 & 0.0368 & 0.0061 & 0.0227 & 0.998 & 0.334 & 0.298 & 0.0742 & 0 & 0 & 0.748 & 0.252 \\ 
 \hline\rowcolor{Green!20}
 \cellcolor{Yellow}   (1,4) & 0 & 0 & 0.098 & 0 & 0 & 0 & -0.170 & -0.170 & 0.0223 & 0 & 0 & 0 & 0.864 & 0.136 \\ \rowcolor{Green!20}
 \cellcolor{Yellow}& 0 & 0 & 0.0102 & 0.0193 & 0 & 0.0119 & 0.521 & -0.177 & 0.165 & 0.0231 & 0 & 0 & 0.856 & 0.144 \\ \rowcolor{Green!20}
 \cellcolor{Yellow}& 0 & 0 & 0.0101 & 0 & 0.0029 & 0 & -0.175 & 0.175 & 0.0229 & 0 & 0 & 0 & 0.859 & 0.141 \\ \rowcolor{Green!20}
 \cellcolor{Yellow}& 0 & 0 & 0.0104 & 0.0198 & 0.0030 & 0.0123 & 0.536 & 0.182 & 0.170 & 0.0237 & 0 & 0 & 0.8505 & 0.149 \\
 \hline\rowcolor{Green!20}
 \cellcolor{Yellow}(1,\frac{9}{2}) & 0 & 0 & 0.0032 & 0 & 0 & 0 & -0.0519 & -0.0519 & 0.0022 & 0 & 0 & 0 & 0.955 & 0.0451 \\ \rowcolor{Green!20}
 \cellcolor{Yellow}& 0 & 0 & 0.0033 & 0.0059 & 0 & 0.0037 & 0.159 & -0.0537 & 0.0526 & 0.0023 & 0 & 0 & 0.952 & 0.0476 \\ \rowcolor{Green!20}
 \cellcolor{Yellow}& 0 & 0 & 0.0032 & 0 & 0.0008 & 0 & -0.0532 & 0.0532 & 0.0023 & 0 & 0 & 0 & 0.953 & 0.0469 \\ \rowcolor{Green!20}
 \cellcolor{Yellow}& 0 & 0 & 0.0033 & 0.0061 & 0.0009 & 0.00038 & 0.1635 & 0.0551 & 0.0540 & 0.0023 & 0 & 0 & 0.9505 & 0.0495 \\ 
 \hline\rowcolor{Green!20}
\cellcolor{Yellow}(2,1) & 0 & 0 & 0.346 & 0 & 0 & 0 & -0.748 & -0.748 & 0.295 & 0 & 0 & 0 & 0.577 & 0.423 \\  \rowcolor{Green!20}
 \cellcolor{Yellow}& 0 & 0 & 0.0381 & 0 & 0.0319 & 0 & -0.846 & 0.824 & 0.326 & 0 & 0 & 0 & 0.5077 & 0.492 \\ 
 \hline\rowcolor{Green!20}
\cellcolor{Yellow} (2,\frac{3}{2}) & 0 & 0 & 0.0171 & 0 & 0 & 0 & -0.315 & -0.315 & 0.0699 & 0 & 0 & 0 & 0.771 & 0.228 \\ 
\rowcolor{Green!20}
 \cellcolor{Yellow}& 0 & 0 & 0.0177 & 0.0358 & 0 & 0.0221 & 0.969 & -0.329 & 0.295 & 0.0723 & 0 & 0 & 0.758 & 0.242 \\ 
 \rowcolor{Green!20}
 \cellcolor{Yellow}& 0 & 0 & 0.0187 & 0 & 0.0113 & 0 & -0.350 & 0.349 & 0.0767 & 0 & 0 & 0 & 0.737 & 0.263 \\  
 \hline\rowcolor{Green!20}
 \cellcolor{Yellow}   (2,2) & 0 & 0 & 0.0032 & 0 & 0 & 0 & -0.0519 & -0.0519 & 0.0022 & 0 & 0 & 0 & 0.955 & 0.0451 \\
  \rowcolor{Green!20}
 \cellcolor{Yellow}& 0 & 0 & 0.0033 & 0.0059 & 0 & 0.0037 & 0.159 & -0.0537 & 0.0526 & 0.0023 & 0 & 0 & 0.952 & 0.0476 \\ 
 \rowcolor{Green!20}
 \cellcolor{Yellow}& 0 & 0 & 0.0035 & 0 & 0.0016 & 0 & -0.0570 & 0.0570 & 0.0024 & 0 & 0 & 0 & 0.948 & 0.0521 \\ 
 \rowcolor{Green!20}
 \cellcolor{Yellow}& 0 & 0 & 0.0036 & 0.0065 & 0.0017 & 0.0040 & 0.1756 & 0.0592 & 0.0579 & 0.0025 & 0 & 0 & 0.945 & 0.552 \\
 \hline\rowcolor{Green!20}
 \cellcolor{Yellow}(3,\frac{1}{2}) & 0 & 0 & 0.0346 & 0 & 0 & 0 & -0.748 & -0.748 & 0.295 & 0 & 0 & 0 & 0.577 & 0.423 \\ \rowcolor{Green!20}
\cellcolor{Yellow} & 0 & 0 & 0.0417 & 0 & 0.0440 & 0 & -0.950 & 0.913 & 0.359 & 0 & 0 & 0 & 0.431 & 0.569 \\
 \hline\rowcolor{Green!20}
 \cellcolor{Yellow}   (3,1) & 0 & 0 & 0.0098 & 0 & 0 & 0 & -0.170 & -0.170 & 0.0223 & 0 & 0 & 0 & 0.864 & 0.136 \\ \rowcolor{Green!20}
 \cellcolor{Yellow}& 0 & 0 & 0.0102 & 0.0193 & 0 & 0.119 & 0.521 & -0.177 & 0.165 & 0.0231 & 0 & 0 & 0.856 & 0.144 \\ \rowcolor{Green!20}
 \cellcolor{Yellow}& 0 & 0 & 0.0118 & 0 & 0.0081 & 0 & 0.208 & -0.208 & 0.0270 & 0 & 0 & 0 & 0.819 & 0.181 \\ \rowcolor{Green!20}
 \cellcolor{Yellow}& 0 & 0 & 0.0123 & 0.0237 & 0.0085 & 0.0147 & 0.641 & 0.218 & 0.200 & 0.0281 & 0 & 0 & 0.8062 & 0.194 \\ 
 \hline\rowcolor{Green!20}
 \cellcolor{Yellow}(4,\frac{1}{2}) & 0 & 0 & 0.0171 & 0 & 0 & 0 & -0.315 & -0.315 & 0.0699 & 0 & 0 & 0 & 0.771 & 0.228 \\ 
 \rowcolor{Green!20}
 \cellcolor{Yellow}& 0 & 0 & 0.0177 & 0.0358 & 0 & 0.0221 & 0.969 & -0.329 & 0.290 & 0.0723 & 0 & 0 & 0.758 & 0.242 \\ \rowcolor{Green!20}
 \cellcolor{Yellow}& 0 & 0 & 0.0226 & 0 & 0.0196 & 0 & 0.439 & -0.437 & 0.0931 & 0 & 0 & 0 & 0.647 & 0.353 \\  
 \hline\rowcolor{Green!20}
 \cellcolor{Yellow}(5,\frac{1}{2}) & 0 & 0 & 0.0033 & 0 & 0 & 0 & -0.0519 & -0.0519 & 0.0022 & 0 & 0 & 0 & 0.955 & 0.0451 \\ \rowcolor{Green!20}
 \cellcolor{Yellow}& 0 & 0 & 0.0033 & 0.0059 & 0 & 0.0037 & 0.159 & -0.0537 & 0.0526 & 0.0023 & 0 & 0 & 0.952 & 0.0476 \\ \rowcolor{Green!20}
 \cellcolor{Yellow}& 0 & 0 & 0.0048 & 0 & 0.0035 & 0 & 0.0798 & -0.0793 & 0.0034 & 0 & 0 & 0 & 0.914 & 0.0859 \\ 
 \rowcolor{Green!20}
 \cellcolor{Yellow}& 0 & 0 & 0.0050 & 0.0092 & 0.0037 & 0.0057 & 0.248 & 0.0843 & 0.0809 & 0.0035 & 0 & 0 & 0.9066 & 0.0934 \\ 
\hline
\end{tabular}
\caption{\small \setstretch{1.0} Fixed points and eigenvalues for vector-like fermions in the fundamental representation of $SU_c(3)$,
in the {\tt 321} approximation scheme.  
The last two columns give the values of the ratio $\sigma$ and $\rho$ for $\alpha_2$ or $\alpha_3$ depending on the case (see \ref{LoopsFPbound}).}
\label{FundNNLO}
\end{table}

\subsubsection{Vector-like fermions in higher representations of $SU_c(3)$} 

For the adjoint representation (with $p=q=1$), the search over the same grid of values for $(N_f,\ell, Y)$ (and thus 126,000 further models) 
does not  produce any \FP within the perturbative domain. 
This is true both in the {\tt 210} and in the {\tt 321} approximation scheme. 

\begin{figure}[h]
\includegraphics[width=\textwidth]{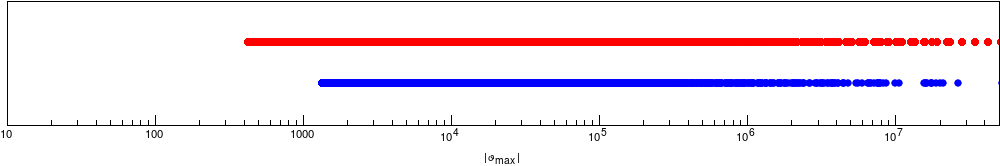}
\caption{\small \setstretch{1.0} Distribution of the largest eigenvalue $\vartheta_{\rm max}$ of the stability matrix of the \FPs of the $SU(3)$ adjoint representation. Blue: eigenvalues for the $Y$-independent solutions. 
Red: eigenvalues for the $Y$-dependent solutions. In both cases, there is no gap and the eigenvalues start at very large values.}
\label{Gapadj}
\end{figure}

In Figure~\ref{Gapadj}, we show the
distribution  the largest eigenvalues of the stability matrix for representative couplings of the fixed points for
the $210$ approximation scheme. We clearly see that the eigenvalues are rather large. In fact, the minimum eigenvalue in the $Y$-independent set
of solutions is $1342$, while in the $Y$-dependent set is $426$.

This problem is confirmed by the study of the central charges. For the $Y$-independent fixed points we find  for all \FPs $\delta a$ of  $O(1000)$. Similarly, for the  $Y$-dependent the fixed points  have   $\delta a$ of  $O(100)$.
Tests of the $c$-function confirm these results, even though the $a$-function seems to be more sensitive, in the sense that it suffers greater relative change. 

Again, we come up empty handed. The models with the vector-like fermions in the adjoint representation of $SU_c(3)$ do not provide a viable AS extension 
to the SM.

Higher $SU_c(3)$ representations are disfavored by experimental constraints because of the early onset of the modifications in the $\alpha_3$ running.

\subsubsection{A model  that almost works}
\label{sec:almost}

Having ruled out all possible candidates, one may wonder if the criteria in  (\ref{def}) and (\ref{thetabound}) might be too stringent and make us miss some potentially interesting models. 
In the case at hand, we can indeed  find additional \FPs that naively seem to be good candidates for an asymptotically 
safe extension of the SM. This is achieved if we allow for larger values of $\vartheta$ and relinquish the condition (\ref{thetabound}). 

As an example, consider the case  with the vector-like fermions in the representations with $N_{f}=3$, $\ell=1/2$ and $Y=3/2$. Its \FPs and eigenvalues 
are given  in Table~\ref{tab:almost}. 

\begin{table}[h]
    \centering
   { \scriptsize \setstretch{1.8} 
    \begin{tabular}{|C|C|C|C|C|C|C|C|C|C|C|C|}\hline
    \rowcolor{Yellow}
  \cellcolor{Yellow}  (N_f, \ell, Y) & \alpha^{*}_{1} & \alpha^{*}_{2} & \alpha^{*}_{3} & \alpha^{*}_{t} & \alpha^{*}_{y} &\vartheta_{1} & \vartheta_{2} & \vartheta_{3} &\vartheta_{4} & \vartheta_{5} & \rho_1 \\ \hline\rowcolor{Green!20}
     \cellcolor{Yellow}      (3,1/2,3/2) & 0.188 & 0  &0 & 0 & 0.778 & 33.2 &-3.36 & -0.817 & 0 & 0  & 2.69 \\
 \hline\rowcolor{Green!20}
    \hline
   \end{tabular}
   }
\caption{\small \setstretch{1.0} Values of the couplings at the fixed point, eigenvalues and $\rho_1$ ratio for the model that almost works ({\tt 210} approximation scheme).}
\label{tab:almost}
\end{table}

This example provides a very interesting (and non-trivial) extension of the SM  which includes non-trivial \FP value for the gauge coupling $\alpha_1$ as well 
as the Yukawa coupling $\alpha_y$ (remember that the quartic scalar interaction in the {\tt 210} scheme does not renormalize).
 
\begin{figure}[h]
    \centering
    \includegraphics[scale=0.7]{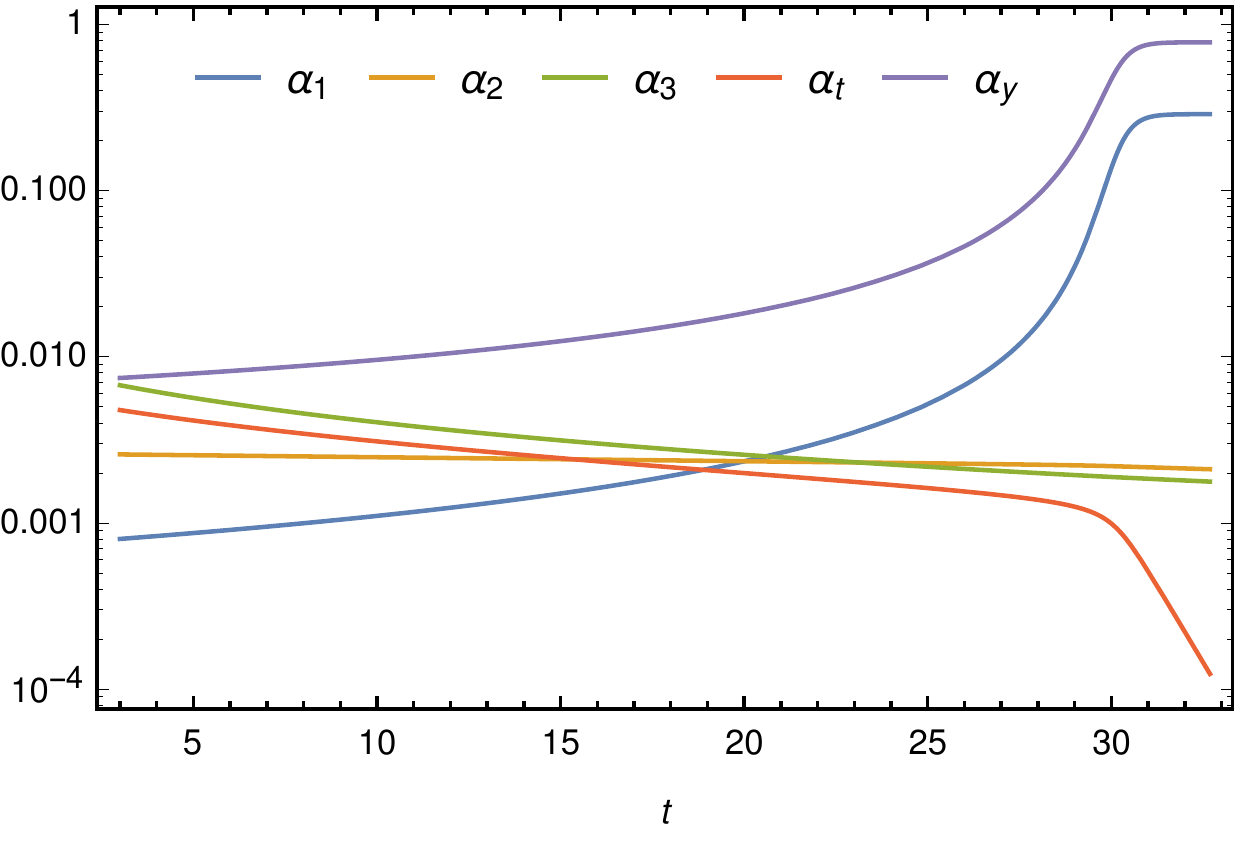}
    \caption{\small \setstretch{1.0} Evolution of the couplings with $t$ in a logarithmic scale for the \FP in Table~\ref{tab:almost}). This running provides a trajectory in the theory space connecting the \FP and the physics at a matching scale around 2 TeV.}
    \label{fig:Matching}
\end{figure}

We  see that some of the scaling exponents $\vartheta_i$ are large and the criterion (\ref{thetabound}) is accordingly violated. 
Nonetheless, let us momentarily suspend disbelief and apply the formula in (\ref{def}). 
We do not find any  coupling frozen to zero and therefore a SM matching seems plausible. 
In fact we find a good matching to the SM couplings in the IR,
with an error of the order of per mille, see Figure~\ref{fig:Matching}.

This model  seems to provide (but for the large scaling exponents) a very promising candidate for an AS extension of the SM. 
Yet  it is not radiatively stable---a fact that vindicates  the role of criteria in (\ref{thetabound}) as a filter for the physical fixed points. The {\tt 321} 
approximation scheme $\beta$-functions generate very different \FPs that cannot be easily traced back to those in the {\tt 210} approximation scheme. 
Moreover, all these fixed points have a trivial coupling
and cannot provide a viable extension to the SM.

\subsubsection{Five benchmark models studied in the literature}
\label{sec:benlit}

The authors of \cite{Litim} find that it is possible to generate asymptotically safe extensions to the SM in the subsystem 
$(\alpha_{2},\alpha_{3},\alpha_{y})$ of the couplings. The five benchmark models discussed in \cite{Litim} (labelled as A, B, C, D and E) are not among those in our scan because 
they do not include hypercharge, top Yukawa and quartic interaction. We analysed them separately. 

The hypercharge $Y$ can easily be added to these models. The 
charge $Y$ must be   larger than a minimal value in order for the corresponding  direction in the UV critical surface to be marginally relevant.  This does not change the
behavior of the models.

Similar to what happens to the model in section \ref{sec:almost}, 
all these models have at least one of their scaling exponents rather large 
(See Tab. \ref{LitimNLO}). The large values of $\vartheta$ imply 
that the \FPs  are not in the perturbative domain even though they can be connected to the SM in the IR regime. The \FPs in the {\tt 210} approximation scheme cannot be connected to those in the {\tt 321} approximation scheme because of their instability against radiative corrections. We can see 
how the structure of the \FPs changes by comparing  Table \ref{LitimNLO} to Table \ref{LitimNNLO}. The eigenvalues are always large in both tables.

\begin{table}[h]
    \centering
   { \scriptsize \setstretch{1.5} 
    \begin{tabular}{|C|C|C|C|C|C|C|C|C|}\hline
    \rowcolor{Yellow}
  \cellcolor{Yellow}  \hspace*{0.7cm} & (R_{3}, R_{2}, N_{f}) & \alpha^{*}_{2} & \alpha^{*}_{3} & \alpha^{*}_{y} & \vartheta_{1} & \vartheta_{2} & \vartheta_{3} & \rho \\ \hline\rowcolor{Green!20}
     \cellcolor{Yellow}  A  & \cellcolor{Yellow} (1,4,12) & 0.241 & 0 & 0.338 & 210 & -1.90 & 0 & 45.3\\
 \hline\rowcolor{Green!20}
  \cellcolor{Yellow} B & \cellcolor{Yellow} (10,1,30) & 0 & 0.129 & 0.116 & 338 & -2.06 & 0 & 107 \\ \rowcolor{Green!20}
 \cellcolor{Yellow} & \cellcolor{Yellow} & 0.277 & 0.129 & 0.116 & 341 & -2.08 & 0.897 & 107\\ 
 \hline\rowcolor{Green!20}
 \cellcolor{Yellow} C & \cellcolor{Yellow} (10,4,80) & 0 & 0.332 & 0.0995 & 23258 & -2.18 & 0 & 9138 \\ \rowcolor{Green!20}
 \cellcolor{Yellow} & \cellcolor{Yellow} & 0.0753 & 0.0503 & 0.0292 & 1499 & 328 & -2.77 & 630 \\ \rowcolor{Green!20}
 \cellcolor{Yellow} & \cellcolor{Yellow} & 0.800 & 0 & 0.150 & 145193 & -2.12 & 0 & 57378 \\ 
 \hline\rowcolor{Green!20}
 \cellcolor{Yellow} D & \cellcolor{Yellow} (3,4,290) & 0.0615 & 0.0416 & 0.0057 & 943 & 45.3 & -2.29 & 371 \\ \rowcolor{Green!20}
 \cellcolor{Yellow} & \cellcolor{Yellow} & 0.0896 & 0 & 0.0067 & 1984 & -2.11 & 0 & 781 \\ 
  \hline\rowcolor{Green!20}
 \cellcolor{Yellow} E & \cellcolor{Yellow} (3,3,72) & 0.218 & 0.150 & 0.0471 & 896 & 112 & -1.78 & 326 \\ 
    \hline
   \end{tabular}
   }
\caption{\small \setstretch{1.0} Couplings, eigenvalues and the ratios $\rho_i$, with $i=2,3$ depending on the case, for the benchmark models in \cite{Litim} for the {\tt 210} approximation scheme.}
\label{LitimNLO}
\end{table}

\begin{table}[h]
    \centering
    \scriptsize \setstretch{1.5}
   \begin{tabular}{|C|C|C|C|C|C|C|C|C|}\hline
    \rowcolor{Yellow}
  \cellcolor{Yellow}  \hspace*{0.7cm}  & (R_{3}, R_{2}, N_{f}) & \alpha^{*}_{2} & \alpha^{*}_{3} & \alpha^{*}_{y} & \vartheta_{1} & \vartheta_{2} & \vartheta_{3} & \rho_3 \\ \hline\rowcolor{Green!20}
     \cellcolor{Yellow} A & \cellcolor{Yellow} (1,4,12) & 0 & 0 & 0.1509 & -4.83 & 0 & 0 & - \\
 \hline\rowcolor{Green!20}
  \cellcolor{Yellow} B & \cellcolor{Yellow} (10,1,30) & 0 & 0.0138 & 0 & -20.02 & 2.24 & 0 & 3.14\\ \rowcolor{Green!20}
 \cellcolor{Yellow} & \cellcolor{Yellow} & 0 & 0 & 0.0594 & -4.75 & 0 & 0 & - \\ 
 \hline\rowcolor{Green!20}
 \cellcolor{Yellow} C & \cellcolor{Yellow} (10,4,80) & 0 & 0 & 0.0187 & -4.501 & 0 & 0 & -\\ \rowcolor{Green!20}
 \cellcolor{Yellow} & \cellcolor{Yellow} & 0 & 0.0036 & 0 & -49.4 & 2.28 & 0 & 9.29 \\ 
 \hline\rowcolor{Green!20}
 \cellcolor{Yellow} D & \cellcolor{Yellow} (3,4,290) & 0 & 0 & 0.0115 & -6.95 & 0 & 0 & - \\ \rowcolor{Green!20}
 \cellcolor{Yellow} & \cellcolor{Yellow} & 0 & 0.0108 & 0 & -36.7 & 1.015 & 0 & 5.81 \\ 
  \hline\rowcolor{Green!20}
 \cellcolor{Yellow} E & \cellcolor{Yellow} (3,3,72) & 0 & 0 & 0.0357 & -5.79 & 0 & 0 & - \\ \rowcolor{Green!20}
 \cellcolor{Yellow} & \cellcolor{Yellow} & 0 & 0.0305 & 0 & -21.8 & 1.098 & 0 & 2.66\\ 
    \hline
   \end{tabular}
\caption{\small \setstretch{1.0} Couplings, eigenvalues and the ratio $\rho_3$ for the benchmark models in \cite{Litim} for the {\tt 321} approximation scheme.}
\label{LitimNNLO}
\end{table}

If we take the \FPs in the {\tt 321} approximation scheme at their face value and try to match them to the SM, we always encounter a coupling, $\alpha_2$ in almost
all the cases (see Table~\ref{LitimNNLO}), that is frozen to its vanishing value: the theory is trivial in the coupling $\alpha_2$ and it cannot be matched to the SM. 
In other words, the benchmark models in \cite{Litim} suffer from the same pathology of  the models in our scan.
Unlike  those models, in this case it is a non-abelian coupling that is trivial. 
This is worrisome and should be born in mind if one were to entertain the idea of embedding $U_Y(1)$ 
in a non-abelian group in order to find an AS extension of the SM.

\subsubsection{Two more  benchmark models studied in the literature}
\label{sec:mann}

The authors of \cite{Mann:2017wzh} study three models where the fermions are in the representations
$(N_f,\, \ell ,\, Y)= (3, 2, 1/6)$, $(3, 1, 0) \oplus (1, 2, 1/2)$ and $(3, 1, 0) \oplus (1, 3, 0) \oplus (1, 1, 1)$, respectively.
Further models (with Majorana fermions charged under only one gauge group at the time) are introduced in \cite{Pelaggi:2017abg}. In both papers, they consider the large $N_f$ limit and
the zeros of the $\beta$-functions are found after resumming the blob diagrams of the perturbative theory. 

Resummed  $\beta$-functions could help in discussing the AS of a theory, yet there is no available procedure that is free of the ambiguities deriving from  summing over a particular class of diagrams. Moreover, a large number of  new states  must be included to be within the regime of validity of the resummation scheme
which---from the phenomenological point of view---is unappealing. 

It is difficult to compare these  results to ours because of the non-perturbative nature of the resummation procedure.  In our approach these models are all ruled out because of the larger scaling dimensions we expect   for the given large values of $N_f$ (see Figure~\ref{fig:EigvalsNTL} and the discussion in section \ref{sec:mod}).  
The \FPs  obtained in the approach of \cite{Mann:2017wzh} and \cite{Pelaggi:2017abg} are probably linked to essential singularities in the complete resummation~\cite{Holdom:2010qs} and no perturbative treatment---like expanding around the \FP values to search for the trajectory back to the SM---is possible. 

While these models provide interesting examples of AS theories, it is difficult to see them as viable candidate for  extensions of the SM because the low-energy matching is problematic: to wit, even assuming that the \FP thus found are physical, the authors of \cite{Pelaggi:2017abg} conclude (in the published version of their paper) that there is no matching because of the persistence of the Landau pole in the $U(1)$ coupling (which can only be avoided at the price of making the vacuum of the model unstable because of the running of the quartic Higgs interaction).

\subsubsection{Combining more than one representation}
\label{sec:more}

Combining vector-like fermions in different representations (as done, for instance, in \cite{Mann:2017wzh,Pelaggi:2017abg}) provides other examples of 
models that almost work.
In the simplest scenario, we can try to construct a model with two types of vector-like fermions. 
In that case, we duplicate the last three terms in Eq. (\ref{Lagrangian}) for fermions 
$\tilde{\psi}$ and scalars $\tilde{S}$.
 We call the extra Yukawa coupling $z$ with, as usual,
\be
\alpha_z=\frac{z^2}{(4\pi)^2}
\ee
and assume no mixing between the two families. 

\begin{table}[h]
    \centering
   { \scriptsize \setstretch{1.8} 
    \begin{tabular}{|C|C|C|C|C|C|C|C|C|C|C|C|C|C|}\hline
    \rowcolor{Yellow}
 \alpha^{*}_{1} & \alpha^{*}_{2} & \alpha^{*}_{3} & \alpha^{*}_{t} & \alpha^{*}_{y} & \alpha^{*}_{z} &
   \vartheta_{1} & \vartheta_{2} & \vartheta_{3} &\vartheta_{4} & \vartheta_{5} &  \vartheta_{6} & \rho_2 & \rho_1 \\ 
   \hline\rowcolor{Green!20}
   0.226 & 0.193  &0 & 0 & 0.778 & 0.534 & 241 & 24.2 & -2.85 & -2.28 & -1.51  &  0 & 0 & 0 \\
 \hline\rowcolor{Green!20}
    \hline
   \end{tabular}
   }
\caption{\small \setstretch{1.0} Values of the couplings at the fixed point of interest, eigenvalues and $\rho$ ratios for the model combining 3 fields in the representation $(1, 2, 3/2)$ and 8 fields in the representation $(1, 5, 0)$ ({\tt 210} approximation scheme).}
\label{tab:FPtwo}
\end{table}

Since many of the BSM extensions attempt to describe dark matter, we take one of the possible minimal models discussed in~\cite{DM} and identify some of the vector-like fermions with dark matter. This exercise makes clear the potential relevance of AS in selecting physics BSM. 

We take $N_{f_2}$ vector-like fermions with quantum numbers $p=q=0$, $\ell=2$ and $Y=0$. That is, we take colorless quintuplets with
no hypercharge. Within the {\tt 210} approximation scheme, for the combination $(1, 2, 3/2) \oplus (1, 5, 0)$, we realize that \FPs split in two categories: 
\FPs that depend on the number of quintuplets $N_{f_2}$ and \FPs that do not. Clearly, the latter have $\alpha_y=0$ so that the vector-like fermions enter only via loops 
in the gauge $\beta$-functions. Consequently, the conditions to lie on the critical surface of those \FPs imply that $\alpha_2=0$. This feature makes the
corresponding \FPs uninteresting.

\begin{figure}[h]
\centering
\includegraphics[scale=0.6]{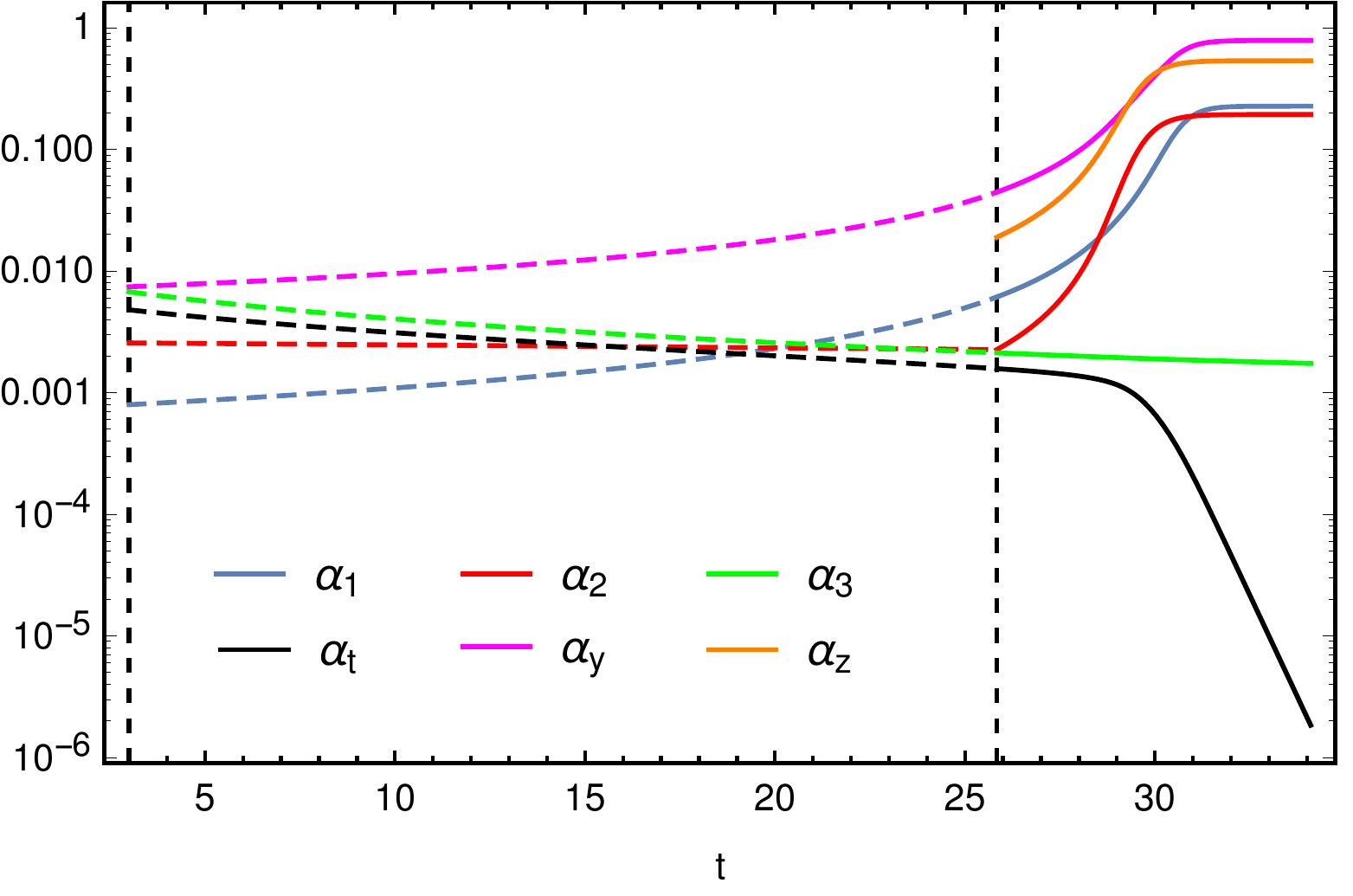}
\caption{\small \setstretch{1.0} Evolution of the couplings with $t$ for the \FP in  Table~\ref{tab:FPtwo} within the {\tt 210} approximation with 3 fields in $(1, 2, 3/2)$ and 8 fields in $(1, 5, 0)$. This running provides a trajectory in the theory space connecting the 
    \FP to a matching scale around $2$ TeV passing through another matching (for the quintuplets) at about $10^{13}$ TeV.}
    \label{fig:Twofam}
\end{figure}

For the remaining fixed points, we find that in order to have $\alpha_i<1$ for all couplings, 
the minimum number of quintuplets should be equal to 
eight. Taking the minimal case of $N_{f_2}=8$, we find $6$ fixed points, all of them having one large eigenvalue around $250$. Thus, according to our requirement
about perturbation theory, these \FPs are not reliable since there is always one $\vartheta$ which is much larger than $1$. This is similar
to what happens in section \ref{sec:almost}. 

Nevertheless, we can  find a matching with the SM. The only difference with respect to the model in section \ref{sec:almost} is that, in the present case, two
matching scales are needed---the reason being that the large number of quintuplets makes $\alpha_2$ decrease fast so that 
these
fields must be decoupled at very high energies. In Figure~\ref{fig:Twofam} we show the logarithmic running of the couplings and the  two different
matching scales. The quintuplets decouple at an energy scale  $O(10^{13})$ TeV (and must be considered {\it wimpzilla} dark matter~\cite{Kolb:1998ki}), 
the doublets at the energy scale of $1.83$ TeV. All the couplings flow to the \FP in Table~\ref{tab:FPtwo}

Even though Figure~\ref{fig:Twofam} shows a nice flow of the coupling constants toward the SM, the size of the eigenvalues spells doom and the likely breakdown of perturbation theory.
Indeed, the \FP analysed does not survive in the {\tt 321} approximation scheme and the model does not work.

\section{Conclusions}

A systematic scan  (covering 378,000 models) of possible extensions of the SM based on vector-like fermions charged under the SM groups and carrying various representations and coming in several copies (generations) shows that there are no \FPs in the $\beta$-function of the models that satisfy the minimal criteria to make them perturbatively stable and therefore physical. Most of those that appear in the {\tt 210} approximation scheme are difficult to identify  when probed in the {\tt 321} approximation scheme. Those that seem to be present in both  schemes (or appear only at the higher order) always  contain a trivial solution in which  at least one of the couplings is frozen to zero thus suggesting that the Landau problem of the LO theory persists at higher orders in the perturbative expansion.

We  conclude that it is not possible, at least with the 378,000 models we have examined, to extend the SM up to arbitrarily high energies in perturbation theory. 
The same happens with the models discussed in section
\ref{sec:almost} and \ref{sec:more}, and  the five models proposed as benchmarks in \cite{Litim}:
 the {\tt 321} approximation scheme $\beta$-functions generate very different \FPs that cannot be easily traced back to those in the {\tt 210} approximation scheme. All the \FPs  in {\tt 321} approximation scheme have  at least one trivial coupling. 
  
Our search for AS extensions of the SM  has returned a negative answer and no viable candidate. This result might indicate that the search must be enlarged to include models with BSM fields more complicated than vector-like fermions. Because the vector-like fermions are actually just a proxy for more general matter fields, this does not seem too promising a line of enquiry. Another possibility is that  the SM gauge groups must be embedded in  larger groups before AS  becomes manifest. All in all, the most plausible scenario seems  that in which the \FPs making the SM AS, if they exist at all,  lay outside the perturbative regime and accordingly are inherently invisible to our approach. Such a non-perturbative computation has been partially explored with negative results in \cite{Pelaggi:2017abg}. Over and above that, we should always bear in mind  the possibility that there exists no AS  extension  to the SM or, if it exists, 
that it requires the inclusion of gravity \cite{Dona:2013qba,Christiansen,Astrid}.

\begin{acknowledgments}
We are very grateful to Luminita Mihaila for giving us access to her results on the 3-loop  $\beta$-functions for the gauge couplings in non-simple groups prior to their publication.
We also thank Daniel Litim and Francesco Sannino for 
useful discussions.
\end{acknowledgments}

\goodbreak

\appendix

\section{Analysis of marginal couplings}
\label{sec:appmarg}

Here we prove the statement, made in Section \ref{sec:marg},
that when the marginal couplings
are associated to vanishing gauge couplings,
the behavior of the flow at quadratic order is determined
by the coefficients $P_{iii}$.
 
The general form of the gauge $\beta$-functions is
\begin{equation}
    \beta_{i}=(A^{i}+B^{i}_{r}\alpha_{r}+C^{i}_{rs}\alpha_{r}\alpha_{s})\,\alpha_{i}^{2},
    \label{GenGauge}
\end{equation}
where $A^{i}$, $B^{i}_{r}$ and $C^{i}_{rs}$ represent the one, two and three-loops coefficients. Their contribution to the stability matrix is given by
\begin{equation}
    M_{ij}=\frac{\partial\beta_{i}}{\partial\alpha_{j}}\biggr|_{\alpha_{i}^{*}}=(B^{i}_{j}+2\,C^{i}_{jr}\alpha_{r}^{*})\, \alpha_{i}^{*2}+2\,(A^{i}+B^{i}_{r}\alpha_{r}^{*}+C^{i}_{rs}\alpha_{r}^{*}\alpha_{s}^{*})\, \alpha_{i}^{*}\delta_{ij}. 
\end{equation}
We see that if $\alpha_{i}^{*}=0$, the row $i$ will have zeros in all the entries. This does not happen for the Yukawa interactions,
whose NLO $\beta$-functions have the form
$\beta_{Y_{i}}=(D^{i}_{r}\alpha_{r}+F^{i}_{rs}\alpha_{r}\alpha_{s})\alpha_{i}$.
Then, the contribution to the stability matrix reads
\begin{equation}
    M_{ij}=\frac{\partial\beta_{Y_{i}}}{\partial\alpha_{j}}\biggr|_{\alpha_{i}^{*}}=(D^{i}_{j}+2F^{i}_{jr}\alpha_{r}^{*})\alpha_{i}^{*}+(D^{i}_{r}\alpha_{r}^{*}+F^{i}_{rs}\alpha_{r}^{*}\alpha_{s}^{*})\delta_{ij}, 
\end{equation}
where we see that if $\alpha_{i}^{*}=0$, the last piece will be in general different from zero. Consequently, we do not have a row of zeros. The fact of having rows of zeros implies that $detM=0$. Thus, the matrix $M$ is singular and there exist vectors $\mathbf{x}$ such that $A\mathbf{x}=0\mathbf{x}$. As a result, $M$ has the eigenvalue $\lambda=0$ with multiplicity given by the number of zero rows. 

Suppose we have a \FP with two gauge couplings equal to zero.
Then the stability matrix will have two zero rows, that
we can assume to be the last two. This implies that the $n-2$ eigenvectors corresponding to
$\lambda_{i}\neq0$ 
have the form $\mathbf{V}^{i}=[V^{i}_{1},V^{i}_{2},\dots,V^{i}_{n-2},0,0]$. The eigenvectors for $\lambda=0$ lie in a $2$-dimensional plane. 
There is a freedom in choosing these vectors, and we can take them to have the form $\mathbf{V}^{n-1}=[V^{n-1}_{1},V^{n-1}_{2},\dots,V^{n-1}_{n-2},V^{n-1}_{n-1},0]$, $\mathbf{V}^{n}=[V^{n}_{1},V^{n}_{2},\dots,V^{n}_{n-2},0,V^{n}_{n}]$. Moreover, the entries $V^{n-1}_{n-1}$, $V^{n}_{n}$ can be taken to be positive without loss of generality. Thus, the transformation matrix constructed with 
the eigenvectors of $M$ takes the form
$\tiny \color{white}
\begin{bmatrix}
a \nn
\end{bmatrix}
$
{\small
\be
{
S = 
\begin{bmatrix} 
    V^{1}_{1} & V^{2}_{1} & \dots & V^{n-2}_{1} & V^{n-1}_{1} & V^{n}_{1} \\
    V^{1}_{2} & V^{2}_{2} & \dots & V^{n-2}_{2} & V^{n-1}_{2} & V^{n}_{2} \\
    \vdots & \vdots & \ddots & \vdots & \vdots & \vdots \\
    V^{1}_{n-2} & V^{2}_{n-2} & \dots & V^{n-2}_{n-2} & V^{n-1}_{n-2} & V^{n}_{n-2} \\
    0 & 0 & \dots & 0 & V^{n-1}_{n-1} & 0 \\
    0 & 0 & \dots & 0 & 0 & V^{n}_{n}
\end{bmatrix}
}
\label{S}
\ee
}
This implies that
{\small
\be
{
    S^{-1}=
\begin{bmatrix}
    a_{1,1} & a_{1,2} & \dots & a_{1,n-2} & a_{1,n-1} & a_{1,n} \\
    a_{2,1} & a_{2,2} & \dots & a_{2,n-2} & a_{2,n-1} & a_{2,n} \\
    \vdots & \vdots & \ddots & \vdots & \vdots & \vdots \\
    a_{n-2,1} & a_{n-2,2} & \dots & a_{n-2,n-2} & a_{n-2,n-1} & a_{n-2,n} \\
    0 & 0 & \dots & 0 & b & 0 \\
    0 & 0 & \dots & 0 & 0 & c
\end{bmatrix}
}
\label{Sinverse}
\ee
}
where we have labelled $a_{i,j}$ the non-zero entries and we have called $b=1/V^{n-1}_{n-1}$, $c=1/V^{n}_{n}$.
Now, when we compute the form of the new variables $z_{i}=S^{-1}_{ij}y_{j}=S^{-1}_{ij}(\alpha_{j}-\alpha_{j}^{*})$, we observe that two of the new coordinates are just proportional to the asymptotically free variables, namely $z_{n-1}=b\cdot y_{n-1}=b\cdot\alpha_{n-1}$, $z_{n}=c\cdot y_{n}=c\cdot\alpha_{n}$. 
This result has an important effect in the analysis.
For the gauge $\beta$-functions, 
\begin{align}
    P_{ijk}=\frac{\partial^{2}\beta_{i}}{\partial\alpha_{j}\alpha_{k}}\biggr|_{\alpha_{i}^{*}}=&2 \, C^{i}_{jk}\alpha_{i}^{*2}+2\, (B^{i}_{j}+2\, C^{i}_{jr}\alpha_{r}^{*})\, \alpha_{i}^{*}\delta_{ik}+2\, (B^{i}_{k}+2C^{i}_{kr}\alpha_{r}^{*})\, \alpha_{i}^{*}\delta_{ij} \\ \notag
    &+2\, (A^{i}+B^{i}_{r}\alpha_{r}^{*}+C^{i}_{rs}\alpha_{r}^{*}\alpha_{s}^{*})\, \delta_{ij}\delta_{ik}
    \label{PGauge}
\end{align}
which in the case of the AF couplings reduces to 
\be
P_{ijk}=2\, (A^{i}+B^{i}_{r}\alpha_{r}^{*}+C^{i}_{rs}\alpha_{r}^{*}\alpha_{s}^{*})\, \delta_{ij}\delta_{ik}\ .
\ee

We conclude that in order to know if a marginal coupling is relevant or irrelevant we need only check the sign of $P_{iii}$. If $P_{iii}<0$, the coupling is marginally relevant. If $P_{iii}>0$, the coupling is marginally irrelevant.

\section{Conformal field theory and central charges}
\label{sec:CFT}

The CFT at a given  \FP  is characterized by two local functions: $c$ and $a$. We refer to them collectively as central charges or CFT functions. They appear in the matrix element of the trace of the energy-momentum tensor of the theory as 
$\langle T_{\mu}^{\mu} \rangle = c W^2 - a E_4 + \cdots $,
where $W$ is the Weyl tensor, $E_4$ is the Euler density, and ellipses denote operators constructed from the fields in the theory.
A function related to the CFT function $a$, often denoted $\tilde{a}$, was proven to be monotonically decreasing following the RG flow from a UV fixed point to an IR one \cite{Osborn,Jack}.
In fact, the RG  flow of the $\tilde{a}$-function is related to the dynamics by means of the $\beta$-functions of the theory; it is given by
\be
\mu \frac{\partial \tilde{a}}{\partial \mu} = - \chi_{ij} \beta^i \beta^j \, ,
\ee
where $\chi_{ij}$ is known as the Zamolodchikov metric.
Evaluated at a fixed point, $\tilde{a}$ reduces to the $a$-function.

In all of the models studied in this paper there is only a UV fixed point present, whereas dynamics in the IR is not known.
Nevertheless, central charges of the UV fixed points can still be used to test whether the fixed points  are reliable.

In any CFT, both $a$ and $c$ have to be positive, and  their ratio has to satisfy the so-called \textit{collider bounds} \cite{Hofman:2008ar}, namely
\be
\frac{1}{3} \leq \left. \frac{a}{c}\right|_{FP} \leq \frac{31}{18} 		\ .
\ee
In perturbation theory, central charges are expanded in series
\bea
\tilde{a}&=&\tilde{a}_{free}+\frac{\tilde{a}^{(1)}}{(4\pi)^2}+\frac{\tilde{a}^{(2)}}{(4\pi)^4}+...	\\
c&=&c_{free}+\frac{c^{(1)}}{(4\pi)^4}+...	\ ,
\eea
and since free-field theory contributions are positive \cite{Duff:1977ay},
\bea
\tilde{a}_{free}&=&\frac{1}{(4\pi)^2} \frac{n_s + 11/2 \, n_w + 62\,n_v}{360}	\\
c_{free}&=&\frac{1}{(4\pi)^2}\frac{1/6\, n_s +n_w + 2\,n_v}{20}
\eea
($n_s$, $n_w$, and $n_v$ referring to scalar, Weyl and vector degrees of freedom,  respectively), the positivity of the CFT functions is ensured in perturbation theory.

There is a  correlation between critical exponents and the change in central charges,
which for the $a$-function can be explained as follows. 
At the fixed point we have,
\be
\tilde{a}^*=a^*=a_{free}+\frac{1}{4}\sum_i b_i \chi_{g_i g_i} \alpha_i^*(1+A_i \alpha_i^*) \,
\ee
where $i$ runs over simple gauge groups, $b_1=B_1, b_2=-B_2, b_3=-B_3$ are the one-loop coefficients of the gauge beta functions, and $\chi_{g_i g_i}$ and $A_i$ are components of the Zamolodchikov metric, see \cite{Dondi:2017civ}. 
One-loop critical exponent follows from $\beta_i=\pm B_i \alpha_i^2$ ($+$ for the group $U(1)$, $-$ otherwise), and reads $\theta^{1L}=2b_i \alpha_i^*$.
Then,
\be
\delta a=\frac{a^*-a_{free}}{a_{free}}=\frac{1}{8a_{free}}\sum_i \theta_i^{1L} \chi_{g_i g_i} (1+A_i \alpha_i^*) \ ,
\ee
which explains the correlation. 

\goodbreak

\section{All the fixed points in the {\tt 210} approximation scheme}
\label{sec:appbeta}

In  Table~\ref{tab:grand} we list all the distinct zeroes of the $\beta$-functions
in the {\tt 210} approximation scheme for all the models discussed in the text and for the SM.
There are altogether 32 zeroes, with the Gaussian fixed point 
appearing with multiplicity four (this is the reason for missing fixed point
$P_{20}$, $P_{27}$, $P_{32}$, which are copies of $P_1$).

\begin{table}[]
\centering
{\scriptsize  \setstretch{2.0} 
\begin{tabular}{|C|C|C|C|C|C|C|}
\hline\rowcolor{Yellow}
 & \alpha^{*}_{1} & \alpha^{*}_{2} & \alpha^{*}_{3} & \alpha^{*}_{t} & \alpha^{*}_{y} & N_f=0  \\ \hline
\cellcolor{Yellow} P_1 & 0 & 0 & 0 & 0 & 0 & (0,0,0,0)  \\
\cellcolor{Yellow} P_2 & 0 & \alpha_2^*(p,q,\ell) & \alpha_3^*(p,q,\ell) & 0 & \alpha_y^*(p,q,\ell) & \left(0,\frac{499}{617},-\frac{319}{2468},0\right)  \\
\cellcolor{Yellow} P_3 & 0 & \alpha_2^*(p,q,\ell) & \alpha_3^*(p,q,\ell) & \alpha_t^*(p,q,\ell) & \alpha_y^*(p,q,\ell) & \left(0,\frac{1226}{1411},-\frac{189}{1411},\frac{277}{1411}\right) \\
\cellcolor{Yellow} P_4 & 0 & \alpha_2^*(p,q,\ell) & \alpha_3^*(p,q,\ell) & 0 & 0 & \left(0,\frac{499}{617},-\frac{319}{2468},0\right) \\ 
\cellcolor{Yellow} P_5 & 0 &\alpha_2^*(p,q,\ell) & \alpha_3^*(p,q,\ell)  & \alpha_t^*(p,q,\ell) & 0 & \left(0,\frac{1226}{1411},-\frac{189}{1411},\frac{277}{1411}\right)  \\
\cellcolor{Yellow} P_6 & \alpha_1^*(p,q,\ell,Y) & \alpha_2^*(p,q,\ell,Y) & \alpha_3^*(p,q,\ell,Y) & 0 & \alpha_y^*(p,q,\ell,Y) & \left(-\frac{7938}{9257},\frac{9841}{9257},-\frac{5395}{37028},0\right)  \\ 
\cellcolor{Yellow} P_7 & \alpha_1^*(p,q,\ell,Y) & \alpha_2^*(p,q,\ell,Y) & \alpha_3^*(p,q,\ell,Y) & \alpha_t^*(p,q,\ell,Y) & \alpha_y^*(p,q,\ell,Y) & \left(-\frac{121821}{142153},\frac{151229}{142153},-\frac{41441}{284306},\frac{427}{142153}\right)  \\ 
\rowcolor{Orange!20}
\cellcolor{Yellow} P_8 & 0 & 0 & \alpha_3^*(p,q,\ell) & \alpha_t^*(p,q,\ell) & \alpha_y^*(p,q,\ell) & \left(0,0,-\frac{9}{38},-\frac{8}{19}\right)  \\
\rowcolor{Orange!20}
\cellcolor{Yellow} P_9 & 0 & 0 & \alpha_3^*(p,q,\ell) & 0 & \alpha_y^*(p,q,\ell) & \left(0,0,-\frac{7}{26},0\right)\\
\rowcolor{Orange!20}
\cellcolor{Yellow} P_{10}& 0 & 0 & \alpha_3^*(p,q,\ell) & \alpha_t^*(p,q,\ell) & 0 & \left(0,0,-\frac{9}{38},-\frac{8}{19}\right) \\
\rowcolor{Orange!20}
\cellcolor{Yellow} P_{11}& 0 & 0 & \alpha_3^*(p,q,\ell) & 0 & 0 & \left(0,0,-\frac{7}{26},0\right)  \\
\cellcolor{Yellow} P_{12}& \alpha_1^*(p,q,\ell,Y) & 0 & \alpha_3^*(p,q,\ell,Y) & 0 & \alpha_y^*(p,q,\ell,Y) & \left(-\frac{225}{943},0,-\frac{1079}{3772}\right)  \\
\cellcolor{Yellow} P_{13}& \alpha_1^*(p,q,\ell,Y) & 0 & \alpha_3^*(p,q,\ell,Y) & \alpha_t^*(p,q,\ell,Y) & \alpha_y^*(p,q,\ell,Y) & \left(-\frac{7266}{16847},0,-\frac{4286}{16847},-\frac{9907}{16847}\right)  \\
\cellcolor{Yellow} P_{14}& \alpha_1^*(p,q,\ell,Y) & 0 & \alpha_3^*(p,q,\ell,Y) & 0 & 0 & \left(-\frac{225}{943},0,-\frac{1079}{3772}\right)  \\
\cellcolor{Yellow} P_{15}& \alpha_1^*(p,q,\ell,Y) & 0 & \alpha_3^*(p,q,\ell,Y) & \alpha_t^*(p,q,\ell,Y) & 0 & \left(-\frac{7266}{16847},0,-\frac{4286}{16847},-\frac{9907}{16847}\right)  \\
\rowcolor{Orange!20}
\cellcolor{Yellow} P_{16}& 0 & \alpha_2^*(p,q,\ell) & 0 & 0 & \alpha_y^*(p,q,\ell) & \left(0,\frac{19}{35},0,0\right)  \\
\rowcolor{Orange!20}
\cellcolor{Yellow} P_{17}& 0 & \alpha_2^*(p,q,\ell) & 0 & \alpha_t^*(p,q,\ell) & \alpha_y^*(p,q,\ell) & \left(0,\frac{38}{61},0,\frac{19}{61}\right)  \\
\rowcolor{Orange!20}
\cellcolor{Yellow} P_{18}& 0 & \alpha_2^*(p,q,\ell) & 0 & 0 & 0 & \left(0,\frac{19}{35},0,0\right)  \\
\rowcolor{Orange!20}
\cellcolor{Yellow} P_{19}& 0 & \alpha_2^*(p,q,\ell) & 0 & \alpha_t^*(p,q,\ell) & 0 & \left(0,\frac{38}{61},0,\frac{19}{61}\right)  \\
\cellcolor{Yellow} P_{21} & \alpha_1^*(p,q,\ell,Y) & 0 & 0 & 0 & 0 & \left(-\frac{123}{199},0,0,0\right)  \\ 
\cellcolor{Yellow} P_{22} & \alpha_1^*(p,q,\ell,Y) & 0 & 0 & \alpha_t^*(p,q,\ell,Y) & 0 & \left(-\frac{2214}{3293},0,0,-\frac{697}{3293}\right)  \\ 
\cellcolor{Yellow} P_{23} & \alpha_1^*(p,q,\ell,Y) & 0 & 0 & \alpha_t^*(p,q,\ell,Y) & \alpha_y^*(p,q,\ell,Y) & \left(-\frac{2214}{3293},0,0,-\frac{697}{3293}\right) \\ 
\cellcolor{Yellow} P_{24} & \alpha_1^*(p,q,\ell,Y) & 0 & 0 & 0 & \alpha_y^*(p,q,\ell,Y) &  \left(-\frac{123}{199},0,0,0\right)  \\ 
\cellcolor{Yellow} P_{25} & \alpha_1^*(p,q,\ell,Y) & \alpha_2^*(p,q,\ell,Y) & 0 & 0 & \alpha_y^*(p,q,\ell,Y) & \left(-\frac{1461}{1559},\frac{1222}{1559},0,0\right)  \\ 
\cellcolor{Yellow} P_{26} & \alpha_1^*(p,q,\ell,Y) & \alpha_2^*(p,q,\ell,Y) & 0 & \alpha_t^*(p,q,\ell,Y) & \alpha_y^*(p,q,\ell,Y) & \left(-\frac{21627}{23569},\frac{515}{637},0,\frac{2719}{23569}\right)  \\ 
\cellcolor{Yellow} P_{28}& \alpha_1^*(p,q,\ell,Y) & \alpha_2^*(p,q,\ell,Y) & 0 & 0 & 0 & \left(-\frac{1461}{1559},\frac{1222}{1559},0,0\right)  \\ 
\cellcolor{Yellow} P_{29}& \alpha_1^*(p,q,\ell,Y) & \alpha_2^*(p,q,\ell,Y) & 0 & \alpha_t^*(p,q,\ell,Y) & 0 & \left(-\frac{21627}{23569},\frac{515}{637},0,\frac{2719}{23569}\right)  \\ 
\cellcolor{Yellow} P_{30}& \alpha_1^*(p,q,\ell,Y) & \alpha_2^*(p,q,\ell,Y) & \alpha_3^*(p,q,\ell,Y) & 0 & 0 & \left(-\frac{7938}{9257},\frac{9841}{9257},-\frac{5395}{37028},0\right)  \\ 
\cellcolor{Yellow} P_{31}& \alpha_1^*(p,q,\ell,Y) & \alpha_2^*(p,q,\ell,Y) & \alpha_3^*(p,q,\ell,Y) & \alpha_t^*(p,q,\ell,Y) & 0 & \left(-\frac{121821}{142153},\frac{151229}{142153},-\frac{41441}{284306},\frac{427}{142153}\right)  \\ 
\hline
\end{tabular}
}
\caption{\small \setstretch{1.0} List of all the \FPs in  the {\tt 210} approximation scheme. When non-zero, the dependence on the quantum numbers is indicated.
Only the highlighted \FPs appear in the tables in the main text.
The column $N_f=0$ contains the values for the SM.}
\label{tab:grand}
\end{table}

The column labelled by $N_f=0$ contains the values of $\alpha_1^*$, $\alpha_2^*$, $\alpha_3^*$, 
$\alpha_t^*$ for the matter content of the SM (the coupling $\alpha_y^*$
does not appear in the SM).
In this case the \FPs all come in pairs. When $N_f\not=0$ this degeneracy
is lifted and all the \FPs are different.

Note that the \FPs  can be roughly divided in two classes.
The \FPs with $\alpha_1^*=0$ have coordinates $\alpha_i^*$ independent of $Y$.
The remaining \FPs have coordinates that in general depend on all the
quantum numbers.

\section{Coefficients of the NLO and NNLO $\beta$-functions}
\label{sec:appcoff}

The $\beta$-function in \eqs{BetaoneII}{Betalambda} contain a number of coefficients that we collect in this appendix.
 The BSM fermions enter in the running of $\alpha_{t}$ via the coefficients
\begin{equation}
    B_{t1}=Y^{2}N_{f}d_{R_{2}}d_{R_{3}}, \ B_{t2}=S_{R_{2}}N_{f}d_{R_{3}}, \ B_{t3}=S_{R_{3}}N_{f}d_{R_{2}}.
\end{equation}
For the BSM Yukawa coupling, besides the terms in Eq. (\ref{YukawaI}), we have the  coefficients
\begin{align} 
        &V=\frac{1}{2}N_{f}^{2}+3\, N_{f}d_{R_{2}}d_{R_{3}}, & &V_{1}=2\, (8\, N_{f}+5\, d_{R_{2}}d_{R_{3}})Y^{2},  \nn \\
     &V_{2}=2\, (8\, N_{f}+5\, d_{R_{2}}d_{R_{3}})C_{R_{2}}, & &V_{3}=2\, (8\, N_{f}+5\, d_{R_{2}}d_{R_{3}})C_{R_{3}}, \nn  \\ 
      &W_{1}=\left(\frac{211}{3}-6Y^{2}+\frac{40}{3}Y^{2}N_{f}d_{R_{2}}d_{R_{3}}\right)Y^{2}, & &W_{12}=12\, Y^{2}C_{R_{2}},  \nn \\
     &W_{2}=\left(-\frac{257}{3}-6C_{R_{2}}+\frac{40}{3}N_{f}S_{R_{2}}d_{R_{3}}\right)C_{R_{2}}, &  &W_{23}=12\, C_{R_{2}}C_{R_{3}},  \nn \\
     &W_{3}=\left(-154-6C_{R_{3}}+\frac{40}{3}N_{f}S_{R_{3}}d_{R_{2}}\right)C_{R_{3}} & &W_{13}=12\, Y^{2}C_{R_{3}}.  
\end{align}
The gauge $\beta$-functions get more contributions. These are split in two classes: the Yukawa contributions:
{\small
\begin{align} 
  K_{y1}&=6\, Y^{2}N_{f}^{3}d_{R_{2}}d_{R_{3}}+7Y^{2}N_{f}^{2}d_{R_{2}}^{2}d_{R_{3}}^{2}, & K_{11}&=6\, Y^{4}N_{f}^{2}d_{R_{2}}d_{R_{3}},  \nn \\
     K_{12}&=6\, Y^{2}C_{R_{2}}N_{f}^{2}d_{R_{2}}d_{R_{3}}, & K_{13}&=6\, Y^{2}C_{R_{3}}N_{f}^{2}d_{R_{2}}d_{R_{3}}, \nn  \\
     K_{y2}&=2\, C_{R_{2}}N_{f}^{3}d_{R_{2}}d_{R_{3}}+\frac{7}{3}C_{R_{2}}N_{f}^{2}d_{R_{2}}^{2}d_{R_{3}}^{2}, & K_{21}&=2\, Y^{2}C_{R_{2}}N_{f}^{2}d_{R_{2}}d_{R_{3}},  \nn \\
    K_{22}&=16\, C_{R_{2}}N_{f}^{2}d_{R_{2}}d_{R_{3}}+2\, C_{R_{2}}^{2}N_{f}^{2}d_{R_{2}}d_{R_{3}} , & K_{23}&=2\, C_{R_{2}}C_{R_{3}}N_{f}^{2}d_{R_{2}}d_{R_{3}}, \nn  \\
    K_{y3}&=\frac{3}{4}C_{R_{3}}N_{f}^{3}d_{R_{2}}d_{R_{3}}+\frac{7}{8}C_{R_{3}}N_{f}^{2}d_{R_{2}}^{2}d_{R_{3}}^{2}, & K_{31}&=\frac{3}{4}Y^{2}C_{R_{3}}N_{f}^{2}d_{R_{2}}d_{R_{3}}, \nn  \\
    K_{33}&=9\, C_{R_{3}}N_{f}^{2}d_{R_{2}}d_{R_{3}}+\frac{3}{4}C_{R_{3}}^{2}N_{f}^{2}d_{R_{2}}d_{R_{3}} & K_{32}&=\frac{3}{4}C_{R_{2}}C_{R_{3}}N_{f}^{2}d_{R_{2}}d_{R_{3}},
\end{align}
}
and the  gauge contributions, which contain the diagonal terms
{\small
\bea
       M_{11}& =&\frac{388613}{2592}+\frac{4405}{162}N_{f}Y^{2}d_{R_{2}}d_{R_{3}}+\frac{463}{9}N_{f}Y^{4}d_{R_{2}}d_{R_{3}} \nn \\
      & &+4N_{f}Y^{6}d_{R_{2}}d_{R_{3}}+\frac{88}{9}N_{f}^{2}Y^{6}d_{R_{2}}^{2}d_{R_{3}}^{2},  \nn \\
  M_{22} &=&\frac{324953}{864}+\frac{13411}{54}N_{f}S_{R_{2}}d_{R_{3}}+\frac{533}{9}N_{f}C_{R_{2}}S_{R_{2}}d_{R_{3}}-4N_{f}C_{R_{2}}^{2}S_{R_{2}}d_{R_{3}} \nonumber \\
 &&-\frac{632}{27}N_{f}^{2}S_{R_{2}}^{2}d_{R_{3}}^{2}-\frac{88}{9}C_{R_{2}}N_{f}^{2}S_{R_{2}}^{2}d_{R_{3}}^{2},  \nn \\
   M_{33}&=&65+\frac{6242}{9}N_{f}S_{R_{3}}d_{R_{2}}+\frac{322}{3}N_{f}C_{R_{3}}S_{R_{3}}d_{R_{2}}-4N_{f}C_{R_{3}}^{2}S_{R_{3}}d_{R_{2}} \nonumber \\
 & &-\frac{316}{9}N_{f}^{2}S_{R_{3}}^{2}d_{R_{2}}^{2}-\frac{88}{9}C_{R_{3}}N_{f}^{2}S_{R_{3}}^{2}d_{R_{2}}^{2}, 
\eea
}
as well as mixed coefficients
{\small
\bea
     M_{12} &=&\frac{205}{48}-8\, C_{R_{2}}N_{f}Y^{4}d_{R_{2}}d_{R_{3}},\qquad 
      M_{13}= \frac{274}{27}+8\, C_{R_{3}}N_{f}Y^{4}d_{R_{2}}d_{R_{3}},  \nn \\
   M_{21}&= &\frac{291}{16}+32\, Y^{2}N_{f}S_{R_{2}}d_{R_{3}}-8\, Y^{2}C_{R_{2}}N_{f}S_{R_{2}}d_{R_{3}},  \nn \\
      M_{23}&=&78+32\, C_{R_{3}}N_{f}S_{R_{2}}d_{R_{3}}-8\, C_{R_{2}}C_{R_{3}}N_{f}S_{R_{2}}d_{R_{3}} ,  \nn \\
 M_{31}&=&\frac{154}{9}+48\, Y^{2}N_{f}S_{R_{3}}d_{R_{2}}-8\, Y^{2}C_{R_{3}}N_{f}S_{R_{3}}d_{R_{2}},  \nn \\
    M_{32}&=& 42+48\, C_{R_{2}}N_{f}S_{R_{3}}d_{R_{2}}-8\, C_{R_{2}}C_{R_{3}}N_{f}S_{R_{3}}d_{R_{2}}, \nn \\
     G_{23}&=&2+8\, C_{R_{2}}C_{R_{3}}N_{f}Y^{2}d_{R_{2}}d_{R_{3}},\quad
     G_{13}=\frac{2}{3}+8\, Y^{2}C_{R_{3}}N_{f}S_{R_{2}}d_{R_{3}}, \nn \\
     G_{12}&=&\frac{1}{4}+8\, Y^{2}C_{R_{2}}N_{f}S_{R_{3}}d_{R_{2}},  
\eea 
}
{\small
\bea
      H_{11}&=&\frac{1315}{32}+\frac{245}{9}C_{R_{2}}N_{f}Y^{2}d_{R_{2}}d_{R_{3}}-4\, C_{R_{2}}^{2}N_{f}Y^{2}d_{R_{2}}d_{R_{3}}+\frac{23}{2}N_{f}S_{R_{2}}d_{R_{3}} \nonumber \\
     & &-\frac{88}{9}C_{R_{2}}N_{f}^{2}Y^{2}S_{R_{2}}d_{R_{2}}d_{R_{3}}^{2}, \nn \\
     G_{11}&=&198+\frac{178}{3}C_{R_{3}}N_{f}Y^{2}d_{R_{2}}d_{R_{3}}-4\, C_{R_{3}}^{2}N_{f}Y^{2}d_{R_{2}}d_{R_{3}}-\frac{968}{27}N_{f}S_{R_{3}}d_{R_{2}}\nonumber \\
    & &-\frac{88}{9}C_{R_{3}}N_{f}^{2}Y^{2}S_{R_{3}}d_{R_{2}}^{2}d_{R_{3}},  \nn  \\ 
    H_{22}&=&\frac{5597}{288}+\frac{23}{6}N_{f}Y^{2}d_{R_{2}}d_{R_{3}}+\frac{463}{9}Y^{2}N_{f}S_{R_{2}}d_{R_{3}}+4N_{f}Y^{4}S_{R_{2}}d_{R_{3}}\nonumber \\
   & &+\frac{88}{9}N_{f}^{2}Y^{4}S_{R_{2}}d_{R_{2}}d_{R_{3}}^{2}, \nn \\
    G_{22}&=&162+\frac{178}{3}C_{R_{3}}N_{f}S_{R_{2}}d_{R_{3}}-4\, C_{R_{3}}^{2}N_{f}S_{R_{2}}d_{R_{3}}-\frac{88}{3}N_{f}S_{R_{3}}d_{R_{2}} \nonumber \\
   & &-\frac{88}{9}C_{R_{3}}N_{f}^{2}S_{R_{2}}S_{R_{3}}d_{R_{2}}d_{R_{3}}, \nn \\
    H_{33}&=&\frac{2615}{108}+\frac{121}{27}N_{f}Y^{2}d_{R_{2}}d_{R_{3}}+\frac{463}{9}Y^{2}N_{f}S_{R_{3}}d_{R_{2}}+4N_{f}Y^{4}S_{R_{3}}d_{R_{2}}\nonumber \\
   & &+\frac{88}{9}N_{f}^{2}Y^{4}S_{R_{3}}d_{R_{3}}d_{R_{2}}^{2}, \nn \\
    G_{33}&=&\frac{109}{4}-11N_{f}S_{R_{2}}d_{R_{3}}+\frac{245}{9}C_{R_{2}}N_{f}S_{R_{3}}d_{R_{2}}-4\, C_{R_{2}}^{2}N_{f}S_{R_{3}}d_{R_{2}} \nonumber \\
   & &-\frac{88}{9}C_{R_{2}}N_{f}^{2}S_{R_{2}}S_{R_{3}}d_{R_{2}}d_{R_{3}},
\eea
}


\bibliographystyle{kp.bst}
\bibliography{Refs_all}

\end{document}